\documentclass[reprint,superscriptaddress,twocolumn,10pt]{revtex4-1}

\usepackage{amsmath}
\usepackage{epsfig}
\usepackage{subfigure,mathrsfs}
\usepackage{array}
\usepackage{amssymb}
\usepackage{braket}
\usepackage{float}
\usepackage{lmodern,amssymb}
\usepackage{physics}
\usepackage[dvipsnames]{xcolor}

\newcommand{\beq}[0]{\begin{equation}}
\newcommand{\eeq}[0]{\end{equation}}
\def\be{\begin{equation}}
\def\ee{\end{equation}}
\def\bea{\begin{eqnarray}}
\def\eea{\end{eqnarray}}
\usepackage{hyperref}

\begin{document}

\title{Microscopic quantum generalization of classical Li\'{e}nard oscillators}

\author{Srijan Bhattacharyya}
\affiliation{Indian Institute of Technology Kanpur, 
Kanpur,Uttar Pradesh 208016, India}
\author{Arnab Ghosh}
\email{arnab@iitk.ac.in}
\affiliation{Indian Institute of Technology Kanpur, 
Kanpur,Uttar Pradesh 208016, India}
\author{{Deb Shankar Ray}}
\affiliation{Indian Association for the Cultivation of Science, Jadavpur, Kolkata 700032, India}

\begin{abstract}
Based on a system-reservoir model and an appropriate choice of nonlinear coupling, we have explored the microscopic quantum generalization of classical Li\'{e}nard systems. Making use of oscillator coherent states and canonical thermal distributions of the associated c-numbers, we have derived the quantum Langevin equation of the reduced system which admits of single or multiple limit cycles. It has been shown that detailed balance in the form of fluctuation-dissipation relation preserves the dynamical stability of the attractors even in case of vacuum excitation. The quantum versions of Rayleigh, Van der Pol and several other variants of Li\'{e}nard oscillators are derived as special cases in our theoretical scheme within a mean-field description.
\end{abstract}

\maketitle

\section{INTRODUCTION}

\par Dissipation is an intriguing issue in physical sciences~\cite{lindenberg_1990_nonequilibrium,*weiss_1999_quantum}. While its incorporation in dynamical problems from the classical point of view is largely phenomenological, the approach turns out to be untenable because of its violation of uncertainty principle, when the system is quantized. The problem is circumvented by coupling the system to a reservoir with infinite degrees of freedom kept at a finite temperature, which allows the fluctuations of the reservoir to act on the system inducing its dissipation or decay~\cite{louisell_1973_quantum, *carmichael_1999_statistical_1}. To maintain thermal equilibrium of the system in contact with reservoir, fluctuation and dissipation get connected through fluctuation-dissipation theorem~\cite{Zwanzig_1973_Nonlinear}. The system-reservoir model describing a dissipative quantum system lies at the heart of the problems related to macroscopic quantum coherence~\cite{Calderia_1981_Influence,
*Calderia_1983_quantum,*Calderia_1987_dynamics}, quantized tunneling in condensed matter physics~\cite{nitzan_2013_chemical}, relaxation processes in quantum optics~\cite{agarwalbookquantumoptics} and magnetic resonance spectroscopy~\cite{munowitz_1988_coherence}, to name a few. An important note in this context is that the dissipative term appearing in the reduced stochastic equation of motion for the system is by and large linear, arising out of linear coupling between the system and the reservoir~\cite{hanggi_1990_reaction, *hernandez_1984_quantal, vladar_1986_theory, *calderia_1993_dissipative, *guinea_1985_bosonization, ferrer_2007_dynamical, *villares_2006_optical}. Introduction of nonlinearity in the dissipative term in a classical system without any noise, on the other hand, may lead to a complete modification of the dynamics. Such nonlinearity in dissipation may result in a force which acts on the system as an intrinsic source of creation of limit cycle~\cite{strogatz_2014_nonlinear}, an asymptotically isolated trajectory in phase space. Li\'{e}nard equation represents a prototypical paradigm for such a classical nonlinear dissipative system of which two special cases are Van der Pol~\cite{b_van_1920, *b_van_1922} and Rayleigh oscillators~\cite{strutt_1877_theory}. The object of the present paper is the search for a quantum analog of such classical Li\'{e}nard oscillator within the framework of system-reservoir theory.

\par Li\'{e}nard system is widely used in describing many oscillating circuits in the development of radio and vacuum tube technology \cite{strogatz_2014_nonlinear,b_van_1920, *b_van_1922, strutt_1877_theory}. The second order differential equation describing classical Li\'{e}nard system is
\be\label{Leinard-eq}
\ddot{x}+f(x)\dot{x}+g(x)=0.
\ee
If $f(x)$ and $g(x)$ follow the Li\'{e}nard's theorem, then the system has at least one unique, stable limit cycle around its origin in the phase plane~\cite{strogatz_2014_nonlinear}. The theorem assures that the \textit{odd function} $g(x)$ acts as a restoring force that tries to reduce any displacement. While assumptions on the \textit{even function} $f(x)$ indicate that it acts as a negative damping at small displacement and positive damping on large displacement. As a result, small oscillations are forced up while large ones are damped down. Thus, it is not difficult to anticipate that the system will settle into a self-sustained oscillation of some intermediate amplitude. Apart from mono-rhythmic models with van der Pol or Rayleigh oscillators, Li\'{e}nard equation depending on the form of the polynomial $f(x)$, may admit bi-rhythmic solutions. The equation has been generalized further to Li\'{e}nard-Smith-Levinson oscillator form that includes the multiple limit cycles~\cite{levinson1942general,saha_2019_reduction,saha_2020_systematic}.

\par The aim of our present work is to propose a microscopic quantum description of  the classical Li\'{e}nard oscillator in the system-reservoir model and subsequent generalization to its several variants. To capture the nature of Li\'{e}nard system, we have defined the interaction Hamiltonian containing appropriate nonlinear coupling terms. The time evolution of the dynamics is followed by Heisenberg equation of motion to obtain operator Langevin equation for the reduced system. The noise and the nonlinear dissipation originating from the system-reservoir coupling are shown to follow the fluctuation-dissipation relation. Our approach is based on a c-number Langevin equation using harmonic oscillator coherent states and canonical thermal distribution of the associated c-numbers for the reservoirs. This ensures that the overall system is thermodynamically  closed. The c-number Langevin equation plays a key role in describing the noisy quantum Li\'{e}nard system. Van der Pol and Rayleigh oscillators are also depicted with a quantum noise term as model examples. Finally our proposal is generalized for several model cases of Eq.~\eqref{Leinard-eq}.

\par The present work is organized as follows: In Sec.~\ref{Sec-II} we introduce the model Hamiltonian to derive reduced dynamics for the system in terms of operator Langevin equation. In the next section oscillator coherent states are used to construct quantum  Langevin equation with c-number noise. Microscopic realization of classical Van der Pol and Rayleigh oscillator are discussed in Sec.~\ref{Sec-IV}. Later on further generalization of the theoretical scheme for arbitrary form of classical Li\'{e}nard systems are presented. Finally we conclude in Sec.~\ref{Sec-V}.

\section{QUANTUM ANALOG OF CLASSICAL LI\'{E}NARD OSCILLATOR}\label{Sec-II}

\par Our search for microscopic description of classical Li\'{e}nard oscillation is based on quantum harmonic oscillator, whose Hilbert space is given by Fock states $|n\rangle$, where $n$ is the number of quanta in that state. The form of the total system-reservoir Hamiltonian is given by
\be \label{model-hamiltonian}
\hat{H} =\hbar\omega_0\hat{a}^\dag\hat{a} +
\sum_{k}\hbar \omega_k \hat{n}_k+i
\hbar \sum_{k}g_{k}[(\hat{a}^\dag)^{n+1}\hat{b}_k-\hat{a}^{n+1}\hat{b}_k^\dag],
\ee
where the first term is the unperturbed system Hamiltonian $\hat{H}_S=\hbar\omega_0\hat{a}^\dag\hat{a}$, the second term represents the free reservoir (bath) Hamiltonian $\hat{H}_R=\sum_{j}\hbar \omega_j \hat{b}_j^\dag\hat{b}_j$, consisting of large number of harmonic oscillators, and the last term describes the interaction Hamiltonian for $n\in {\mathbb{Z}}^+$. Here $\hat{a}$ and $\hat{b}_k$ are the annihilation operators for system and reservoir respectively which fulfil the following commutation relations:
\be\label{commutation-reln}
[\hat{a},\hat{a}^\dag]=1,\quad [\hat{b}_m,\hat{b}_n^\dag]=\delta_{mn},\quad [\hat{a},\hat{b}_k]=0. 
\ee
The elementary exchange of energy between the system and the reservoir consists of single quantum absorption from the $k$-th bath mode  and simultaneous creation of $(n+1)$ quanta of excitation in the system mode and vice-versa.

\par Since our object here is to recover the dissipative dynamics of the Li\'{e}nard oscillation from the system-reservoir Hamiltonain Eq.~\eqref{model-hamiltonian} for which dissipation is always accompanied by an internal quantum noise, we expect a modification for Eq.~\eqref{Leinard-eq} in the following form in a c-number description;
\be\label{noisy-Leinard-eq}
\ddot{x}+f(x)\dot{x}+\omega^{2}_{0}x=\eta(t).
\ee
Here $\eta(t)$ is a Gaussian white $\delta$-correlated quantum noise with zero mean. In other words, we look for a connection between the nonlinear dissipation $f(x)\dot{x}$ and stochastic noise term $\eta(t)$ which allows the dynamical system to admit stable but noisy limit cycle oscillations.

\par From Eq.~\eqref{model-hamiltonian}, using the commutation relations [Eq.~\eqref{commutation-reln}], we can easily evaluate the Heisenberg operator equations for the system and bath degrees of freedom as
\be\label{eom-a(t)}
\dot{\hat{a}}(t)=-i\omega_0\hat{a}(t) +(n+1)\sum_{k}g_k ({\hat{a}^\dag})^n\hat{b}_k(t),
\ee
and 
\be\label{eom-bj(t)}
\dot{\hat{b}}_j (t)=-i\omega_j \hat{b}_j (t)-g_j \hat{a}^{n+1} (t),
\ee
respectively. Formally integrating Eq.~\eqref{eom-bj(t)}, we get
\be\label{int-eom-bj(t)}
 \hat{b}_j (t) =\hat{b}_j e^{-i\omega_jt} -g_j \int_{0}^t \hat{a}^{n+1} (t^\prime) e^{-i\omega_j (t-t^\prime )} dt^\prime,
\ee
where the first term is the free evolution of the bath operator, while the second term is arising due to the interaction with the system. Inserting Eq.~\eqref{int-eom-bj(t)} into Eq.~\eqref{eom-a(t)}, we find
\begin{eqnarray}\label{eff-eom-a(t)}
  \dot{\hat{a}}(t)=-i\omega_0 \hat{a}(t) +(n+1)\sum_{k} g_k \hat{b}_{k} (\hat{a}^\dag)^n (t) e^{-i\omega_k t}\nonumber\\ 
  - (n+1)\sum_{k} g_k^2 (\hat{a}^\dag)^n (t) \int_{0}^t \hat{a}^{n+1}(t^\prime) e^{-i\omega_k (t- t^\prime)} dt^\prime.
\end{eqnarray}
Now introducing slowly varying operator $\hat{A}(t)=\hat{a}(t)e^{i\omega_0 t}$ in Eq.~\eqref{eff-eom-a(t)}, which varies little over the inverse reservoir bandwidth, we can take the system operator out of the integral by substituting $\hat{A}(t^\prime) \simeq \hat{A}(t)$, under Markov approximation~\cite{louisell_1973_quantum,
carmichael_1999_statistical_1}. Then replacing the remaining integral of $t^\prime$ by the usual $\delta$-function~\cite{agarwalbookquantumoptics}, we finally arrive at the reduced operator equation for the system which is given by
\begin{gather}\label{op-langevin-eq}
\dot{\hat{A}}(t)=-\gamma_{n+1}(\hat{A}^\dag)^n (t) \hat{A}^{n+1}(t)
+\hat{F}_{n+1}(t)(\hat{A}^\dag)^n (t),
\end{gather}
and its hermitian adjoint. The reduced dynamics described by the operator Langevin Eq.~\eqref{op-langevin-eq}, contains the usual dissipative term 
\be\label{dmaping-coeff}
\gamma_{n+1}= (n+1)\sum_{k}g_k^2 \pi \delta(\omega_k -(n+1)\omega_0),
\ee
as well as the noise term
\be\label{noise-op}
\hat{F}_{n+1}(t)=(n+1)\sum_{k}g_k \hat{b}_k\exp[-i\{\omega_k -(n+1)\omega_0\}t],
\ee
which is multiplicative in nature. The noise operator in Eq.~\eqref{op-langevin-eq} appears as a natural consequence of the system-reservoir coupling. It is thus imperative that the Li\'{e}nard system with a microscopic basis must be internally noisy. Combining Eq.~\eqref{op-langevin-eq} and its hermitian adjoint we may further define a noise operator $\hat{G}_{n+1}(t)=\hat{F}_{n+1}(t)+\hat{F}^{\dag}_{n+1}(t)$.

\par The properties of the reservoir can then be calculated by thermal averaging over appropriately ordered noise operators. To this end we define quantum statistical average of any reservoir operator $\hat{O}$ as
\be\label{qs-avg-O(t)}
\langle \hat{O}(t)\rangle_{qs}=\frac{\Tr[\hat{O}\exp(-{\hat{H}_R}/{KT})]}{\Tr[\exp(-{\hat{H}_R}/{KT})]},
\ee
where $\hat{H}_R=\sum_{j}\hbar \omega_j \hat{n}_j$ at $t=0$ and $\hat{n}_j$ denotes the number operator for the $j$-th bath mode.

\par Based on the above considerations, the noise properties of the operator may be calculated by using the canonical thermal distribution Eq.~\eqref{noise-op}. It can be shown that the noise is zero-centered, so that
\be\label{qs-avg-G_n+1}
\langle \hat{G}_{n+1} (t)\rangle_{qs}=0,
\ee
and satisfies the following relation
\bea\label{Op-FD-reln}
&&\Re\left[\langle \hat{F}_{n+1}^\dag (t) \hat{F}_{n+1}(t^\prime) +\hat{F}_{n+1}(t) \hat{F}_{n+1}^\dag(t^\prime)\rangle_{qs} \right]\nonumber\\
&&=(n+1)^2 \sum_{k} g_k^2 \langle (2\hat{n}_k^B +1)\rangle_{qs} \cos\{{[\omega_k -(n+1)\omega_0](t-t^\prime)}\}\nonumber\\
&&=(n+1)^2 \sum_{k} g_k^2 \coth{\left(\frac{\hbar \omega_k}{2KT}\right)}\cos\{{[\omega_k -(n+1)\omega_0](t-t^\prime)}\}\nonumber\\
\eea
where the cotangent hyperbolic factor in Eq.~\eqref{Op-FD-reln} can be identified with the Bose-Einstein distribution
\be
\langle \hat{n}_k^B\rangle_{qs}=\frac{1}{e^{\hbar \omega_k /KT}-1}=\Bar{n}_B (\omega_k),
\ee
using the following relation
\be\label{qs-avg-2nB+1}
\langle (2\hat{n}_k^B +1)\rangle_{qs}=2\Bar{n}_B (\omega_k) + 1 = \coth{\left(\frac{\hbar \omega_k}{2KT}\right)}.
\ee
Equation~\eqref{Op-FD-reln} refers to the fluctuation-dissipation relation for a bosonic bath, which guarantees that the overall system is thermodynamically closed. In the subsequent sections we show that the above detailed balance helps us to preserve the dynamical stability of the limit cycle even in presence of noise. An external noise of even very weak intensity on the other hand destroys the attractors. The plus one factor in Eq.~\eqref{qs-avg-2nB+1} is responsible for vacuum fluctuation which is always present at quantum scale even at absolute zero. Its implication on the microscopic realization of limit cycles will be analyzed in the following sections.

\section{C-NUMBER DESCRIPTION OF QUANTUM LANGEVIN EQUATION}\label{Sec-III}

\par The main purpose of this section is to construct a quantum Langevin equation with c-number noise. As a first step, we return to Eq.~\eqref{op-langevin-eq} and carry out the quantum mechanical average $\langle ... \rangle$ over the initial product separable quantum states of the system oscillator and the bath oscillators at $t=0$, $\ket{\alpha}\ket{\mu_1}\ket{\mu_2}...\ket{\mu_k}...\ket{\mu_N}$  to obtain
\begin{gather}\label{qm-avg-op-langevin-eq}
\langle \dot{\hat{A}}(t)\rangle=-\gamma_{n+1}\langle(\hat{A}^\dag)^n (t) \hat{A}^{n+1}(t)\rangle
+\langle\hat{F}_{n+1}(t)\rangle\langle(\hat{A}^\dag)^n (t)\rangle.
\end{gather}
Here $\ket{\alpha}$ refers to the initial coherent state of the system and $\{\ket{\mu_k}\}$ correspond to initial coherent states of the bath operators. The rationale behind using harmonic oscillator coherent states for the system and bath operators is to recast Eq.~\eqref{qm-avg-op-langevin-eq} as classical-looking Langevin equation for the reduced system oscillator interacting with a bosonic bath.

\par Defining the following quantum mechanical averages for the system and noise operators 
\begin{gather}
\langle \hat{A}(t)\rangle = \alpha(t); \langle \hat{A}^{\dag}(t)\rangle = \alpha^{*}(t);\;
\langle\hat{G}_{n+1}(t)\rangle = \xi_{n+1} (t),
\end{gather}
Eq.~\eqref{qm-avg-op-langevin-eq} and its adjoint can be written as
\bea\label{c-number-Langevin-eq}
\dot{\alpha}(t) &=& -\gamma_{n+1} |\alpha|^{2n} \alpha + f_{n+1}(t) (\alpha^*)^n,\nonumber\\
\dot{\alpha}^*(t) &=& -\gamma_{n+1} |\alpha|^{2n} \alpha^{*} + f_{n+1}^*(t) \alpha^n.
\eea
Here the c-number quantum noise $\xi_{n+1} (t)$ is given by
\begin{eqnarray}\label{c-number-noise}
\xi_{n+1} (t)&=&f_{n+1} (t) 
+f^{*}_{n+1} (t),\nonumber\\ 
&=& (n+1) \sum_{k} g_k \left[\mu_k (0) \exp\{-i(\omega_k -(n+1)\omega_0)t\}\right.\nonumber\\ 
&&\left.+\;\mu^{*}_k (0) \exp\{i(\omega_k -(n+1)\omega_0)t\}\right],\nonumber\\
\end{eqnarray}
where $\mu_k$ and $\mu_k^*$ are the associated c-numbers for the bath operators. In deriving Eqs.~\eqref{c-number-Langevin-eq} from Eq.~\eqref{qm-avg-op-langevin-eq}, we only consider normal ordering for the system operators. Instead for use of different ordering of the operators, we may end up with altogether different (but equivalent~\cite{louisell_1973_quantum}) forms for the nonlinear damping. We will return to this point in the next section when we discuss the quantum-classical correspondance for the limit cycles within the framework of the same interaction Hamiltonian under mean-field approximation~\cite{landaubook,*callenbook, Owen_2018_quantum, navarrete_2017_general}.

\par Now to realize $\xi_{n+1} (t)$ as an effective c-number noise, we introduce the ansatz that $\mu_k (0)$ and $\mu_k^* (0)$ in Eq.~\eqref{c-number-noise} are distributed according to Wigner thermal canonical distribution of Gaussian form~\cite{hilley_1984_distribution} as follows
\be\label{Wigner-dist}
W_k^B [\mu_k (0),\mu_k^* (0)]=N_B \exp{-\frac{|\mu_k (0)| ^2}{2 \coth{\left(\frac{\hbar \omega_k}{2KT}\right)}}}. 
\ee
Here $N_B$ is the normalization constant and $\coth{\left(\frac{\hbar \omega_k}{2KT}\right)}$ is the width of the distribution. For any arbitrary quantum mechanical mean value of a bath operator $ \langle \hat{B}_k \rangle $, which is a function of $\mu_k (0)$ and $\mu_k^* (0)$, its statistical average can then be calculated as
\be\label{statistical-avg-def}
\langle\langle\hat{B}_k \rangle\rangle_{s}=\int   \langle\hat{B}_k \rangle  W_k^B [\mu_k (0),\mu_k^* (0)]d\mu_k (0) d\mu_k^* (0).
\ee
Using the ansatz of Eq.~\eqref{Wigner-dist} and definition of statistical average Eq.~\eqref{statistical-avg-def}, one can show that the c-number noise $\xi_{n+1} (t)$ satisfies the following relations, respectively,
\be\label{eta_n+1-c-no-avg}
\langle \xi_{n+1} (t)\rangle_{s}=0,
\ee
and
\begin{eqnarray}\label{c-no-FD-reln}
\lefteqn{\langle \xi_{n+1}(t) \xi^{*}_{n+1} ({t^\prime})\rangle_{s}}\nonumber\\
&=&(n+1)^2 \sum_{k} g^{2}_{k}\coth{\left(\frac{\hbar \omega_k}{2KT}\right)}\cos\{(\omega_k-(n+1)\omega_0)(t-{t^\prime})\}.\nonumber\\
\end{eqnarray}
Equations~\eqref{eta_n+1-c-no-avg} and~\eqref{c-no-FD-reln} imply that the c-number noise $\xi_{n+1} (t)$ is zero-centered and follows the fluctuation-dissipation relation as expressed in Eq.~\eqref{Op-FD-reln}. Therefore, Eqs.~\eqref{c-no-FD-reln} and~\eqref{Op-FD-reln} are equivalent. In order to identify the connection between Eqs.~\eqref{Op-FD-reln} and~\eqref{c-no-FD-reln}, it is sometime convenient to write down the fluctuation-dissipation relation in the following form:
\begin{eqnarray}\label{c-no-and-op-FD-reln}
&&\langle \xi_{n+1}(t) \xi^{*}_{n+1} ({t^\prime})\rangle_{s}\nonumber\\
&&=\Re\left[\langle \hat{F}_{n+1}^\dag (t) \hat{F}_{n+1}(t^\prime) + \hat{F}_{n+1} (t) \hat{F}^\dag_{n+1}(t^\prime) \rangle_{qs}\right],\nonumber\\
&&=2(n+1) \gamma_{n+1} \coth{\left[\frac{(n+1)\hbar \omega_0}{2KT}\right]}\delta(t-t^{\prime}).
\end{eqnarray}
In deriving the above expression we have assumed that the bath modes are closely spaced in frequency so that one may replace the summation over $k$ in Eq.~\eqref{c-no-FD-reln} by an integral over $\omega$ using the density of states $\rho(\omega)$~\cite{agarwalbookquantumoptics} which yields $\gamma_{n+1}=(n+1)\pi g^{2}((n+1)\omega_0)\rho((n+1)\omega_0)$. Secondly, Eq.~\eqref{c-no-and-op-FD-reln} depends on the noise operator ordering, but not on time ordering. This indicates a clear-cut advantage of the c-number formalism which allows us to bypass the operator ordering prescription for the derivation of noise properties~\cite{barik_2005_quantum,*ghosh_2012_canonical}. The c-number noise $\xi_{n+1} (t)$ as characterized by Eqs.\eqref{eta_n+1-c-no-avg} and~\eqref{c-no-FD-reln} is classical-looking in form but essentially quantum mechanical in nature. Therefore, the essence of microscopic origin of quantum limit cycles can be captured in the present formalism quite effectively in terms of c-number description by simply implementing the techniques of classical non-equilibrium statistical mechanics. Above formalism has been extensively used in several earlier occasions in connection with quantum optics, chemical dynamics, and multidimensional transition state theory in the context of spin and bosonic baths~\cite{banerjee_2002_approach,barik_2005_quantum,
*sinha2011quantum,*sinha_2011_decay,
ghosh2011dissipation,
*ghosh_2012_canonical,*sinha2013brownian}.

\section{CONSTRUCTION OF QUANTUM LI\'{E}NARD SYSTEM; APPLICATION OF THE PROPOSED MODEL}\label{Sec-IV}

\par We are now in a position to apply our method to various nonlinear systems that produce limit cycle oscillations. Two such classic examples are Van der Pol and Rayleigh oscillators, whose basic equation of motions are 
\be\label{cl-vanderpol-osc}
\ddot{x}+\omega_0^2 x -\epsilon (1-x^2)\dot{x}=0,
\ee
and 
\be\label{cl-rayleigh-osc}
\ddot{x}+\omega_0^2 x -\epsilon (1-\dot{x}^2)\dot{x} =0,
\ee
respectively, where we assume $\epsilon >0$. A broad class of biological and chemical oscillations are modelled either in terms of Rayleigh or Van der Pol oscillator or in terms of their generalization ~\cite{goldbeter_1997_biochemical,
kar_2003_collapse,ghosh_2014_lienard,*ghosh_2015_rayleigh}. According to Eq.\eqref{noisy-Leinard-eq}, both the above models are subject to internal quantum noise satisfying fluctuation-dissipation theorem, when derived from their respective microscopic Hamiltonians. The form of the interaction for these specific examples simplifies to
\be\label{H_I-for-n=1}
\hat{H}_I=i\hbar \sum_{k}g_{k}[(\hat{a}^\dag)^{2}\hat{b}_k-(\hat{a})^{2}\hat{b}_k^\dag],
\ee
which is a special case of Eq.~\eqref{model-hamiltonian}. From this interaction Hamiltonian [Eq.~\eqref{H_I-for-n=1}], we proceed as in Sec. \ref{Sec-II} and finally arrive at the following operator Langevin equations for the reduced system 
\begin{gather}\label{op-langevin-eq:n=1}
\dot{\hat{A}}(t)=-\gamma_{2}\hat{A}^\dag(t) \hat{A}^{2}(t)
+\hat{F}_{2}(t)\hat{A}^\dag (t),
\end{gather}
which is the generalized operator Eq.~\eqref{op-langevin-eq} for $n=1$. If we follow the usual normal ordering prescription discussed in Sec. \ref{Sec-III}, the resulting quantum dynamics becomes identical to the generalized c-number Eq.~\eqref{c-number-Langevin-eq} with $n=1$:
\bea\label{c-number-Langevin-eq:n=1}
\dot{\alpha}(t) &=& -\gamma_{2} |\alpha|^{2} \alpha + f_{2}(t) \alpha^*.
\eea

\par
\begin{figure}[H]
	\centering		\includegraphics[width=1.15\columnwidth]{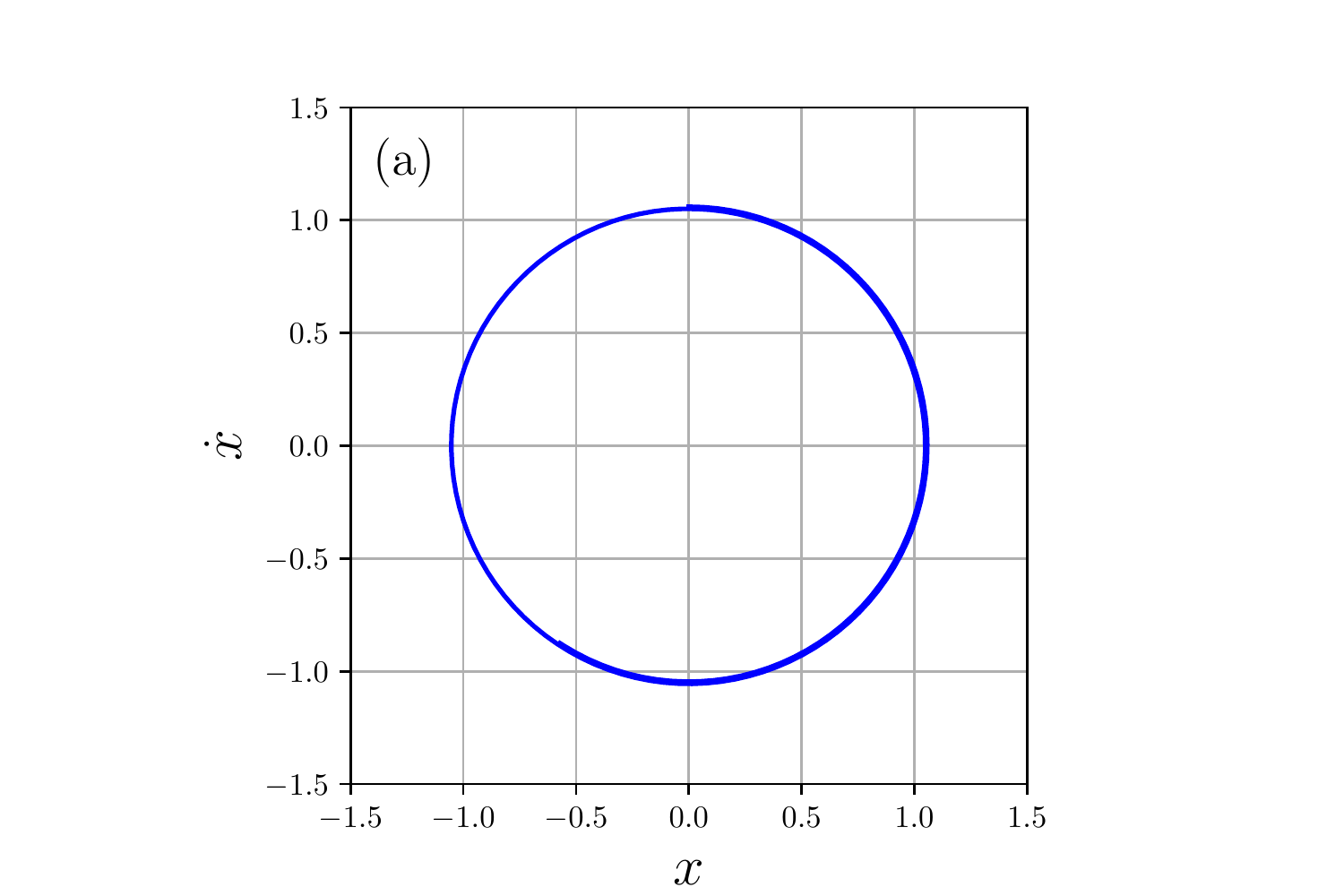}		\includegraphics[width=1.15\columnwidth]{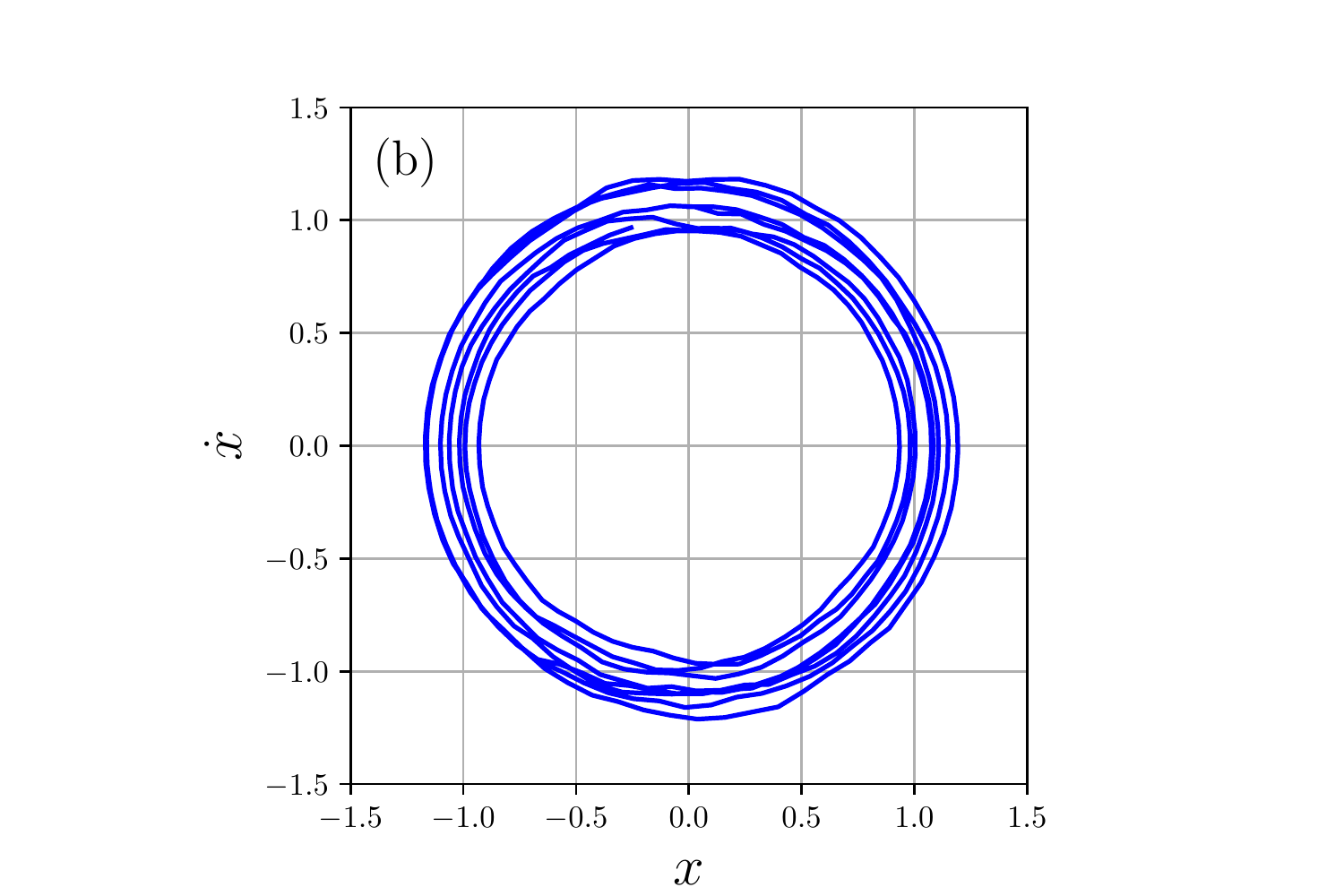}		\includegraphics[width=1.15\columnwidth]{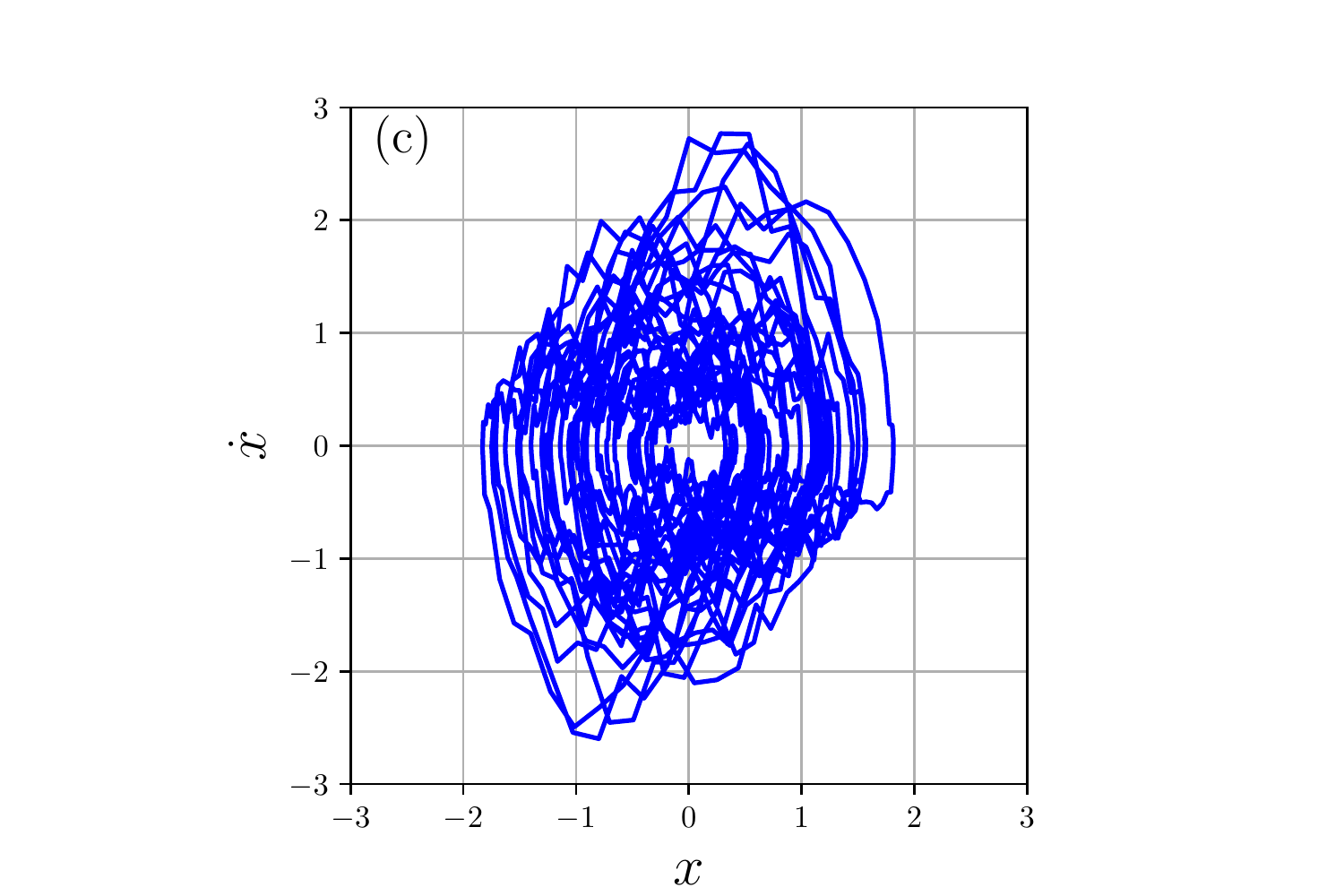}
\caption{The phase space trajectories of (a) classical and (b) quantum Li\'{e}nard oscillator~[Eq.\eqref{Lienard-eq-f(x)=x2}] with $f(x)=\epsilon x^2$ are plotted in absence (a) and presence (b) of intrinsic noise. (c) Quantum Li\'{e}nard oscillator in presence of external noise: lack of detailed balance destroys the limit cycle oscillation. We set $\epsilon =0.01$ and $KT=50$ (units arbitrary).} 
\label{Fig:I} 
\end{figure}

 Substituting $\alpha$ and $\alpha^*$ by 
\begin{eqnarray}\label{substitution-van-der-Pol}
\alpha=\frac{1}{2}\left(x-\frac{i\dot{x}}{\omega_0}\right)\exp[-i\omega_0 t],\nonumber\\
\alpha^{*}=\frac{1}{2}\left(x+\frac{i\dot{x}}{\omega_0}\right)\exp[i\omega_0 t],
\end{eqnarray}
and collecting the terms of right order from Eq.~\eqref{c-number-Langevin-eq:n=1}, we have the following differential equation for the system
\be\label{Lienard-eq-f(x)=x2}
\ddot{x}+\gamma_{2}x^{2} \dot{x}+\omega_0^2 x = \eta_{2} (x,\dot{x}),
\ee 
where 
\begin{eqnarray}
\eta_{2} (x,\dot{x})=
\frac{i\omega_0}{2} \left[f_{2}\left(x+\frac{i\dot{x}}{\omega_0}\right) e^{2i\omega_0 t}\right.\nonumber\\
\left.-f_{2}^*\left(x-\frac{i\dot{x}}{\omega_0}\right) e^{-2i\omega_0 t} \right],
\end{eqnarray}
is the quantum noise term arising out of the system-reservoir interaction.
Equation~\eqref{Lienard-eq-f(x)=x2} may be regarded as the simplest representation of the quantum Li\'{e}nard system [Eq.~\eqref{noisy-Leinard-eq}] where $f(x)= \epsilon x^2$ and $g(x)=\omega_0^2 x$. The typical phase portraits of the system in absence and presence of internal noise are shown in Fig.~\ref{Fig:I}. We use python code for random number generators~\cite{linge_2016_programming}  to simulate the Gaussian white noise~\cite{jung_1990_invariant,*barik_2005_quantum_2}. Throughout our paper we choose $\omega_0=1$ for numerical calculations. In presence of noise, the stochastic trajectories neither collapse to a steady state nor diverge. This implies that the detailed balance in the form of fluctuation-dissipation [Eq.~\eqref{c-no-FD-reln}] guarantees the dynamical stability of motion [Fig.~\ref{Fig:I} b]. If the detailed balance in the form of Eq.~\eqref{c-no-FD-reln} is not properly maintained, the limit cycle is destroyed. This is illustrated in [Fig.~\ref{Fig:I} c] when the noise is of external origin.          

\par Equations~\eqref{op-langevin-eq:n=1} (and~\eqref{c-number-Langevin-eq:n=1}) are the quantum Li\'{e}nard equation in operator (and c-number) form. In order to recover the two special cases of this system we proceed as follows: we first note that the use of commutation relation $[\hat{A},\hat{A}^\dag]=1$ in the operator Eq.~\eqref{op-langevin-eq:n=1} leads to several other equivalent forms~\cite{louisell_1973_quantum}. For example, we have for Eq.~\eqref{op-langevin-eq:n=1}
\begin{gather}\label{op-langevin-eq:n=1,use:commt}
\dot{\hat{A}}(t)=-\gamma_{2}(\hat{A}\hat{A}^\dag \hat{A}-\hat{A})
+\hat{F}_{2}(t)\hat{A}^\dag (t).
\end{gather}

More generally we may write for Eq.~\eqref{op-langevin-eq}
\begin{eqnarray}\label{op-langevin-eq:gen, use:commt}
\dot{\hat{A}}(t)&=&-\gamma_{n+1}\sum_{p,q...r}\phi(p,q..r) \hat{A}^{p}(\hat{A}^\dag)^{q}... \hat{A}^{r} \nonumber \\
&&+\hat{F}_{n+1}(t)(\hat{A}^\dag)^{n} (t).
\end{eqnarray}
In the next step we derive the c-number equivalent of the above operator equation such that it corresponds to the specific classical Li\'{e}nard form. This may be achieved by appropiate operator ordering followed by performing average with coherent states under mean field approximation. We are then led a desired form of c-number equivalent of Eq.~\eqref{op-langevin-eq:gen, use:commt}

\begin{gather}\label{c-no-langevin-eq:gen, use:commt}
\dot{\alpha}= -\gamma_{n+1}\sum_{i,j} \psi(i,j)\alpha^{i} (\alpha^*)^{j} +f_{n+1}(t)(\alpha^*)^n.
\end{gather}

\par A pertinent point is to be noted. Recognizing the right hand side of Eq.~\eqref{c-no-langevin-eq:gen, use:commt} as a polynomial in c-numbers, the term with its highest power is \textit{sufficient} for the existance of limit cycle as shown in Eq.~\eqref{c-number-Langevin-eq:n=1}. The inclusion of terms of lower power is \textit{necessary} to retain the shape of the limit cycle that conforms faithfully with its classical counterpart. The mean field approximation therefore allows us to establish a quantum-classical correspondence. Furthermore we emphasize that the stability of the limit cycles and their number remain invariant with respect to the inclusion of terms with lower power of the polynomial. Based on these considerations Eq.~\eqref{op-langevin-eq:n=1,use:commt} takes the following form (For details see Appendix)

\begin{eqnarray}\label{c-no-Langevin-eq-gen:n=1}
\dot{\alpha}
&=& -\frac{\gamma_2}{m_1} (m_1 |\alpha|^2 + m_0)\alpha + f_2 \alpha^*,
\end{eqnarray}
which may correspond to the specific Li\'{e}nard system for specific values of $m_1$ and $m_0$. In what follows, we will show that Van der Pol and Rayleigh oscillators appear from it as special cases.

\subsection{Van der Pol oscillator}

\par The Van der Pol oscillator is a prototypical self-sustained oscillator which has been used to model the dynamics of a variety of classical~\cite{pikovskybook} and biological processes, such as heart~\cite{b_van_1928}, neurons~\cite{fitzhugh_1961}, and circadian rhythms~\cite{jewett_1998_refinement}. If we substitute $m_1=1$ and $m_0=-1$ in Eq.~\eqref{c-no-Langevin-eq-gen:n=1}, we find 
\bea\label{noisy-van-der-Pol}
\dot{\alpha}(t) &=& -\gamma_{2} (|\alpha|^{2}-1) \alpha + f_{2}(t) \alpha^*,
\eea
and its complex conjugate, where the c-number noise expression follows from Eq.~\eqref{c-number-noise} as
\be 
f_2 =2 \sum_k g_k \mu_k (0)\exp[-i(\omega_k -2\omega_0)t].
\ee
If we further replace $\alpha$ and $\alpha^*$ in Eq.~\eqref{noisy-van-der-Pol} by the usual form [Eq.~\eqref{substitution-van-der-Pol}] and neglect the high frequency terms, one may obtain the following differential equation after little bit of rearrangement
\be\label{eom-van-der-Pol-osc}	
\ddot{x}+\gamma_2 (x^2-1)\dot{x}+\omega_0^2 x=\eta_2 (x,\dot{x}),
\ee
where by definition $\gamma_2 >0$.

\begin{figure}[H]
	\centering
		\includegraphics[width=1.15\columnwidth]{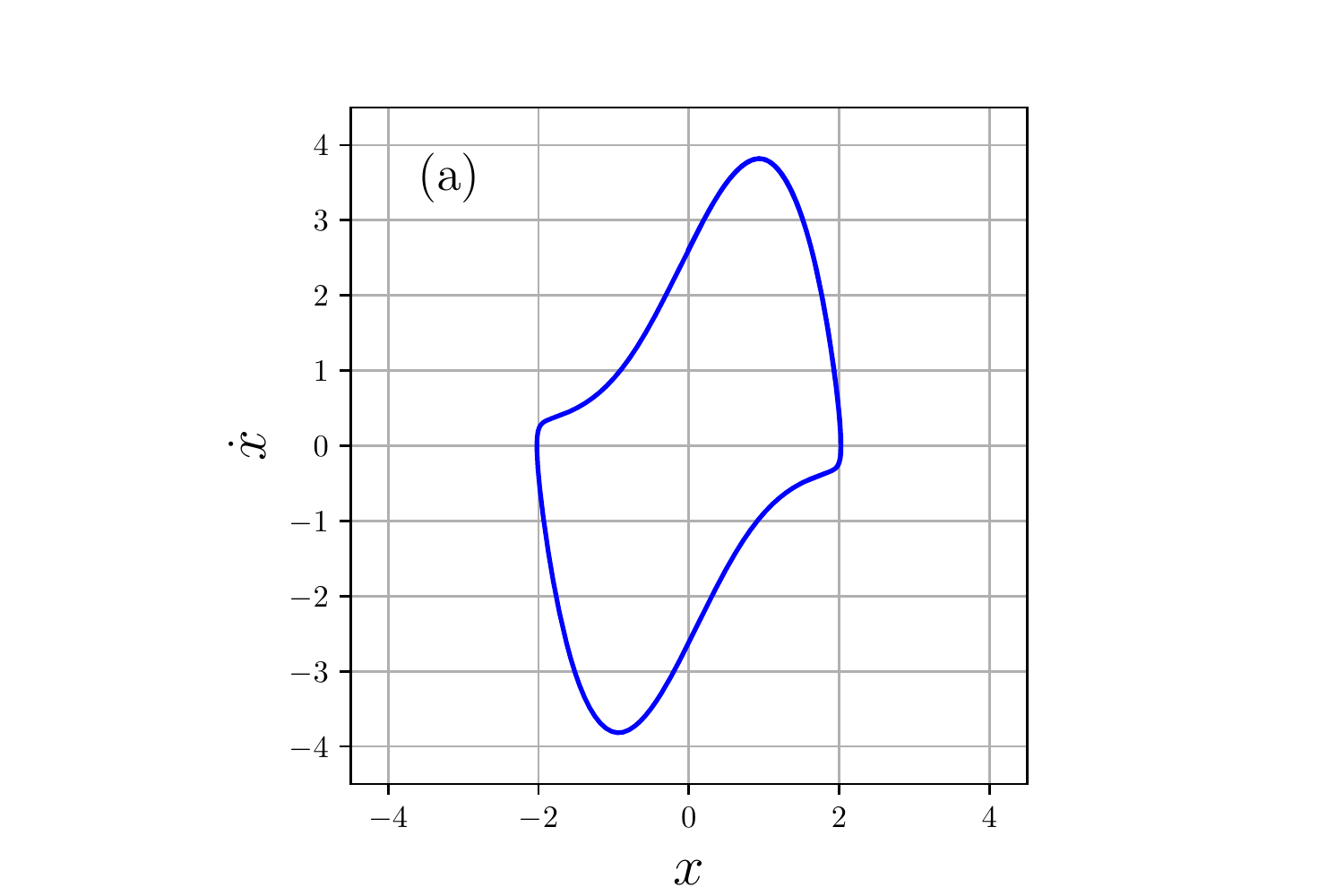} 
		\includegraphics[width=1.15\columnwidth]{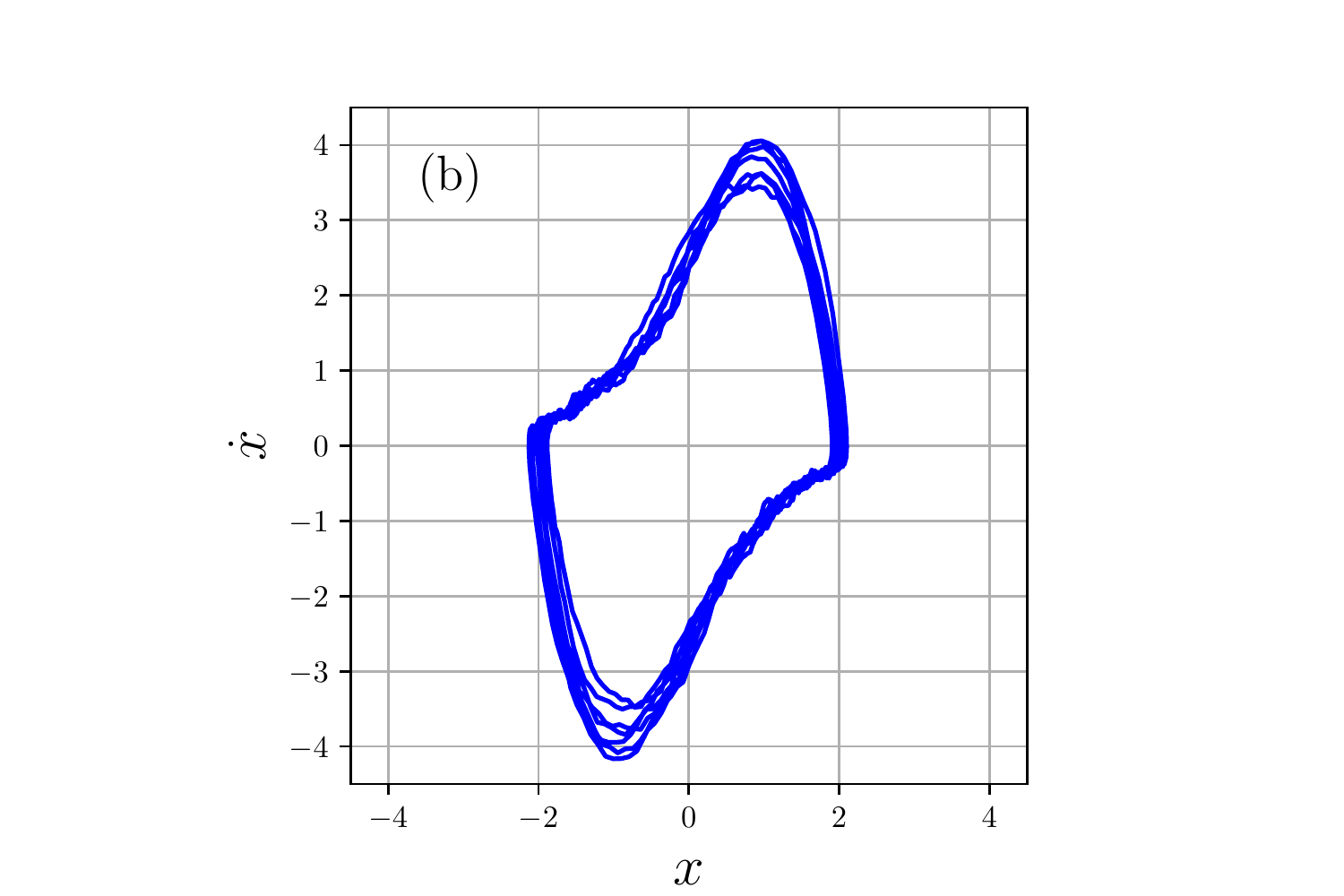}
\caption{(a) The phase space trajectories of a (a) classical [Eq.~\eqref{cl-vanderpol-osc}] and (b) quantum Van der Pol oscillator [Eq.~\eqref{eom-van-der-Pol-osc}] are plotted for $\epsilon=2$ and $KT=0$. It is shown that quantum fluctuations can retain the general shape of the limit cycle oscillation even at absolute zero (units arbitrary).} 
\label{Fig.2}   
\end{figure}

 The c-number Eq.~\eqref{eom-van-der-Pol-osc} looks similar to classical equation of Van der Pol oscillator which is simultaneously driven by the c-number quantum noise. This quantum Van der Pol oscillator can therefore be regarded as a quantum harmonic oscillator with two types of dissipation: negative $(-\dot{x})$ as well as nonlinear $(x^2 \dot{x})$ damping, and an internal noise --- combination of which leads to a noisy but self-sustained limit cycle oscillation. The presence of limit cycle suggests that for the system which follows the isolated asymptotic trajectory, the gain in energy due to self-excitation is equal to the loss of energy due to dissipation. The fluctuation-dissipation relation on the other hand entails the detailed balance condition which implies that the energy dissipated by the system on an average is equal to the energy gained by it through fluctuation of the reservoir. The energy loss-gain in the former processes is purely dynamical while that as a result of detailed balance is statistical in nature.

\par Because of its simple form, such quantum Van der Pol oscillator has recently gained significant interest for realization of synchronization phenomena at quantum scale ~\cite{lee_2013_quantum,Ishibashi2017oscillation,walter_2014_quantum,lee_2014_entanglement}. Interestingly, experiment with trapped-ions may serve as an ideal test bed for simulating collective dynamics for such oscillators ~\cite{leibfried_2003_quantum,haffner_2008_quantum,blatt_2012_quantum,monroe_2013_scaling}. The phase space diagrams for both classical van der Pol oscillator and its quantum counterpart are plotted in Fig.~\ref{Fig.2} for illustration.

\subsection{Rayleigh oscillator}

\par Similar to Van der Pol oscillator, Rayleigh oscillator can also be obtained from same interaction Hamiltonian Eq.~\eqref{H_I-for-n=1} or in other words from the same generalized c-number Langevin dynamics [Eq.~\eqref{c-no-Langevin-eq-gen:n=1}]. This immediately suggests that both the noisy Van der Pol and Rayleigh oscillators are equivalent to the simplest possible form of Li\'{e}nard system [Eq.~\eqref{c-number-Langevin-eq:n=1}] within a mean-field description. Thus the characteristic features of both the two oscillators can be essentially captured by Eq.~\eqref{Lienard-eq-f(x)=x2}.

\par To recover the explicit form of Rayleigh oscillator, we substitute $m_1=3$ and $m_0=-1$ in Eq.~\eqref{c-no-Langevin-eq-gen:n=1}, so that we get
\be \label{c-no-Langevin-eq-rayleigh-osc}
\dot{\alpha}(t)=\frac{\gamma_2}{3}(1-3|\alpha|^2)\alpha + f_2  \alpha^*,
\ee
and its complex conjugate. To arrive at the standard form of the Rayleigh oscillator equation from Eq.~\eqref{c-no-Langevin-eq-rayleigh-osc}, instead of Eq.~\eqref{substitution-van-der-Pol}, we now use the following substitution for $\alpha$ and $\alpha^*$,
\begin{eqnarray}\label{substitution-rayleigh-osc}
\alpha=\frac{1}{2}(\omega_0 x-i\dot{x})\exp[-i\omega_0 t],\nonumber\\ 
\alpha^*=\frac{1}{2}(\omega_0 x+i\dot{x})\exp[+i\omega_0 t]. 
\end{eqnarray}
This difference in the substitution can be traced back to the dimensional relationship between the phase space dynamical variables of the two oscillators as understandable from Eqs.~\eqref{cl-vanderpol-osc} and~\eqref{cl-rayleigh-osc}. With the help of Eqs.~\eqref{substitution-rayleigh-osc} and using the same treatment as before, we finally reduce the following equation of motion for the quantum Rayleigh oscillator 
\be\label{noisy-Rayleigh-osc}
\ddot{x}+\frac{1}{3}\gamma_2(\dot{x}^2-1)\dot{x}+\omega_0^2 x=\eta_2 (x,\dot{x}).
\ee

\begin{figure}[H]
	\centering
		\includegraphics[width=1.15\columnwidth]{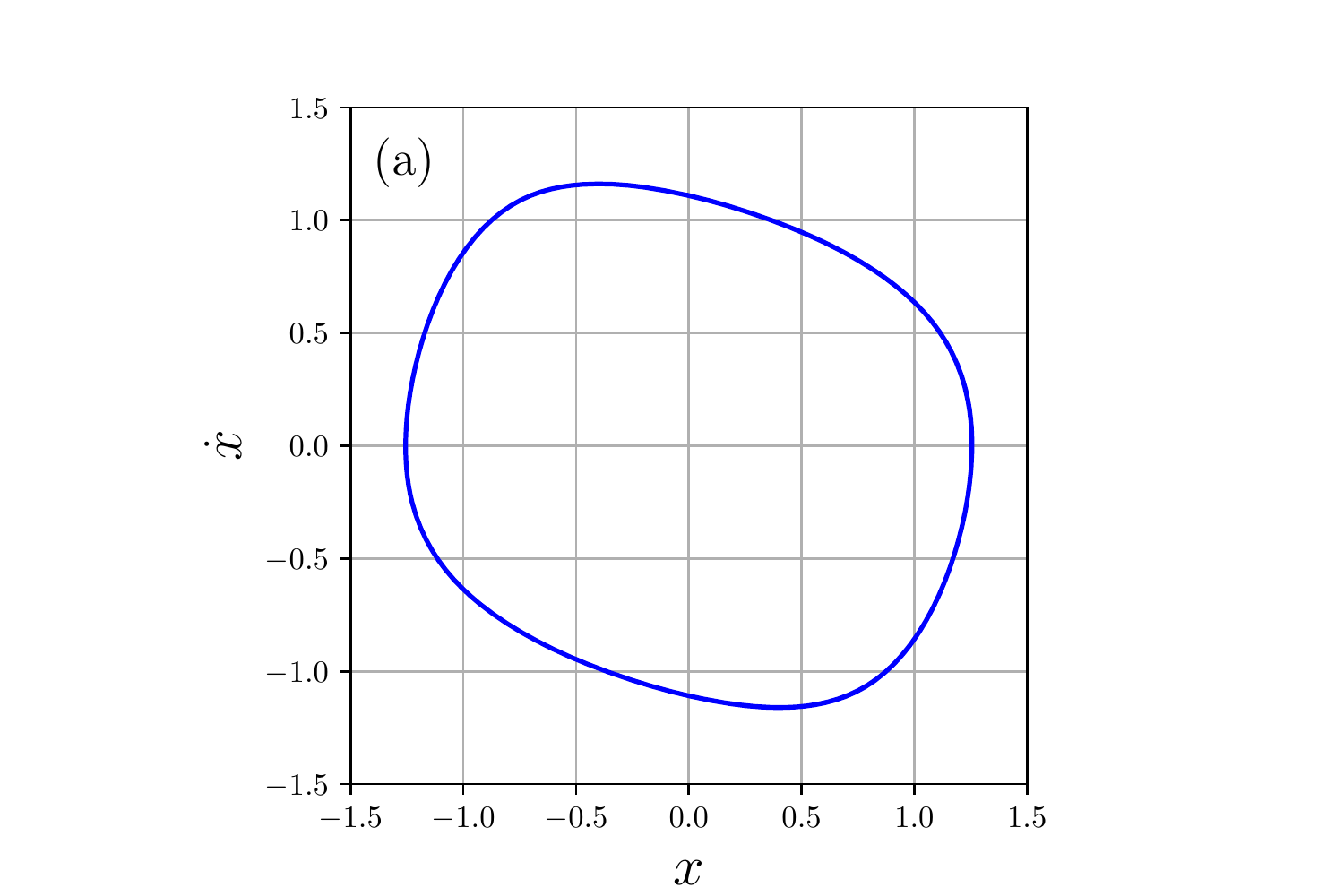}
		\includegraphics[width=1.15\columnwidth]{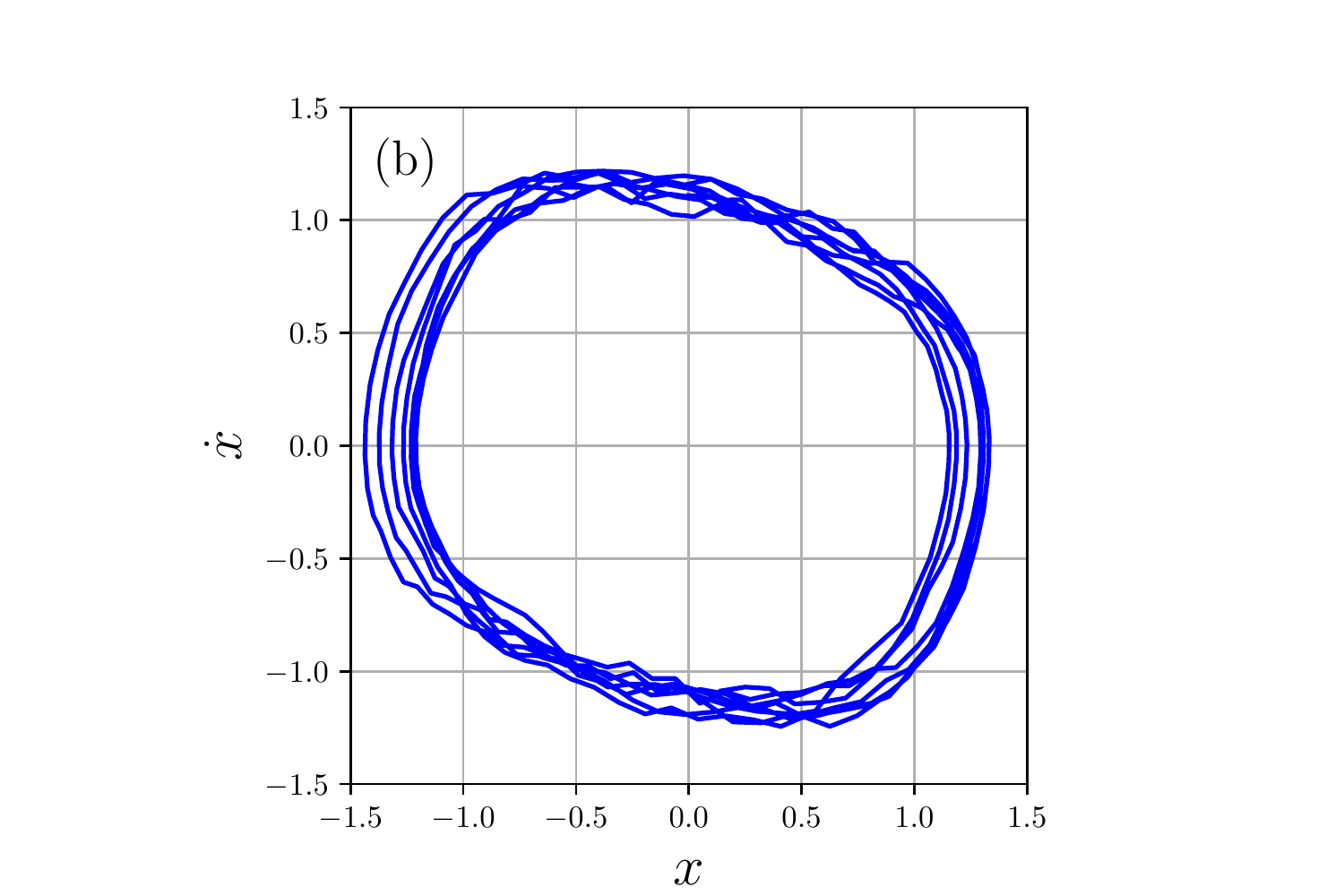}
\caption{(a) The phase space trajectories of a (a) classical [Eq.~\eqref{cl-rayleigh-osc}] and (b) quantum Rayleigh oscillator [Eq.~\eqref{noisy-Rayleigh-osc}] are plotted for the parameters $\epsilon=1$ and $KT=10$ (units arbitrary). } \label{Fig.III} 
\end{figure}

Similar to Van-der Pol oscillator, Rayleigh oscillator plays a close kinship with parametric oscillation in pipe organs, as pioneered by Rayleigh~\cite{lord_1883_mainatained} and in wave mixing in nonlinear optics~\cite{yariv_2000_introduction}. The phase space plots for classical Rayleigh oscillator and its quantum version are depicted in Fig.~\ref{Fig.III}. In general, we find that average quantum dynamics agrees with the phase boundary of the classical limit cycle even at the microscopic scale.

\subsection{Construction of generalized quantum Li\'{e}nard oscillators}

Let us now extend our formalism to construct a broader class of Li\'{e}nard oscillators. For this we adopt the following recipe: First we resort to operator Langevin Eq.~\eqref{op-langevin-eq} and discuss the first two cases [$n=1$ and $2$ of Eq.~\eqref{op-langevin-eq}] as illustrative examples. Next we generalize the concept(see Appendix for details) to construct the most generic form of the Li\'{e}nard oscillators that can be derived from our model Hamiltonian Eq.~\eqref{model-hamiltonian}.

$\bullet$ \textbf{Case-1: $f(x)=\epsilon(a_1 x^2 +a_0)$:---} In the previous section we have shown that  a general form of $f(x)=\epsilon(a_1 x^2 +a_0)$ can be obtained from the interaction Hamiltonian $\hat{H}_I=i\hbar \sum_{k}g_k [(\hat{a}^\dag)^2 \hat{b}_k -(\hat{a})^2\hat{b}_k^\dag]$. The respective c-number Langevin dynamics is given by Eq.~\eqref{c-no-Langevin-eq-gen:n=1} and the corresponding equation of motion takes the form of   
$\ddot{x}+\omega_0^2 x=-(\gamma_2/m_1)(m_1 x^2 + m_0)\dot{x}+\eta_2(x,\dot{x})$, where we may choose $m_1$ and $m_0$ so that it is consistent with the values $a_1$, $a_0$ and $\epsilon$ of $f(x)$. Essentially one sets $\epsilon=(\gamma_2/m_1)$, then $m_1$ and $m_0$ become identical to $a_1$ and $a_0$ respectively. This can be regarded as a representation of Li\'{e}nard system in quantum scenario with the friction coefficient $f(x)=\epsilon(a_1 x^2 +a_0)$, such that our desired system exhibits a single quantum limit cycle.

$\bullet$ \textbf{Case-2: $f(x)=\epsilon (a_2 x^4+a_1 x^2 +a_0)$:---} Let us now consider most general form of $f(x)$ with the next highest order power in accordance with the Li\'{e}nard's theorem. In the similar spirit one can show that the above form of $f(x)$ can be obtained from the Hamiltonian
\be
\hat{H}_I=i\hbar \sum_{k}g_k [(\hat{a}^\dag)^3 \hat{b}_k -(\hat{a})^3\hat{b}_k^\dag ],
\ee
which corresponds to $n=2$ of the generalized interaction Hamiltonian [Eq.~\eqref{model-hamiltonian}]. From the operator Langevin equations Eq.~\eqref{op-langevin-eq}, one carry out the same procedure to derive the c-number Langevin equations for $n=2$ (see Appendix)
\be\label{c-no-Langevin-eq:n=2}
\dot{\alpha}=-\frac{\gamma_3}{m_2}\left[m_2|\alpha|^4 + m_1 |\alpha|^2 +m_0\right] \alpha+f_3(\alpha^*)^2.
\ee
Following the standard substitution as given in Eq.~\eqref{substitution-van-der-Pol} we find
\be
\ddot{x}+\omega_0^2 x=-\frac{\gamma_3}{m_2}\left[\frac{m_2}{2} x^4 + m_1 x^2 + m_0\right] \dot{x}+\eta_3 (x,\dot{x}).
\ee
\begin{table}[H]
 \caption
  \centering
\begin{center}\begin{tabular}{c c c c c}

\hline \hline 
\\
Ex. \quad & \quad $m_2$ \quad & \quad \quad $m_1$ \quad \quad & \quad \quad $m_0$ \quad \quad  & \quad $f(x)$ \\ [1ex]
\hline
\\
$1$ & $2$ & $1$ & $-1$ & $\epsilon(x^4 +x^2 -1)$\\
\\
$2$ & $2$ & $-1$ & $0$ & $\epsilon (x^4-x^2)$ \\
\\
$3$ & $2$ & $0$ & $-1$ & $\epsilon (x^4 -1)$ \\ [1ex]
\hline
\end{tabular}\end{center}
\label{Table-I}
\end{table}
This is the relevant quantum version of Li\'{e}nard system with $f(x)=\epsilon (a_2 x^4+a_1 x^2 +a_0)$, where we have to choose the values of $m_2$, $m_1$ and $m_0$ judiciously, such that it matches with the known form of $f(x)$. Some of the choices of $\{m_2,m_1,m_0\}$ are shown in the Table~\ref{Table-I}. Note that in each case the value of $\epsilon$ varies as $\gamma_3/m_2$. It was shown by Rychkov~\cite{rychkov_1975_maximal} that the number of limit cycles is at most two for $f(x)=\epsilon(a_2x^4+ a_1x^2+ a_0)$. Now extrapolating the form of the above two cases we may generalize the formula by the method of mathematical induction to an arbitrary polynomial $f(x)$ as given below:

$\bullet$ \textbf{Case-3: $f(x)=a_n x^{2n} +a_{n-1}x^{2n-2}...+ a_1 x^2 + a_0$:-}     
Let us now consider the most general form of an even function $f(x)$. From the operator  equation of motion ~\eqref{op-langevin-eq} one can evaluate the following equations for the c-number Langevin dynamics (see Appendix)
\begin{eqnarray}\label{c-no-Langevin-eq:gen}
\dot{\alpha}&=&-\frac{\gamma_{n+1}}{m_n}\left[m_n |\alpha |^{2n} + m_{n-1} |\alpha |^{2n-2} .... +m_1 |\alpha |^2 \right.\nonumber \\
&&\left.+ m_0 \right] \alpha +f_{n+1} (\alpha^*)^n,
\end{eqnarray}

\begin{eqnarray}\label{c-no-Langevin-eq:gen-2}
\dot{\alpha}^*&=&-\frac{\gamma_{n+1}}{m_n}\left[ m_n |\alpha |^{2n} + m_{n-1} |\alpha |^{2n-2} ....+m_1 |\alpha |^2 \right.\nonumber \\
&&\left.+ m_0 \right] \alpha^* +f_{n+1}^* \alpha^n,
\end{eqnarray} 
where $m_0$, $m_1$, .... $m_n$ are independent of one another. Proceeding in the similar way, it is possible to construct the following dynamical equation of motion for the system 
\begin{eqnarray}\label{eom-gen-Leinard-osc}
\ddot{x}+\omega_0^2 x &=&-\frac{\gamma_{n+1}}{m_n}(a_n x^{2n} +a_{n-1} x^{2n-2}....+a_1 x^2 +a_0 ) \dot{x}\nonumber \\
&&+ \eta_{n+1}(x,\dot{x}),
\end{eqnarray}
where 
\begin{equation}
a_j =\frac{m_j}{^{2j}C_{j+1}} j, \quad \forall \quad j=1,2...n \quad \text{and} \quad a_0 =m_0.
\end{equation}
Above equation gives us the relation between different $m_j$'s with $a_j$'s. Note that for $n=1$ (Case-1), $a_j=m_j$ irrespective of all $j$.

\par
\begin{figure}[H]
	\centering
		\includegraphics[width=1.15\columnwidth]{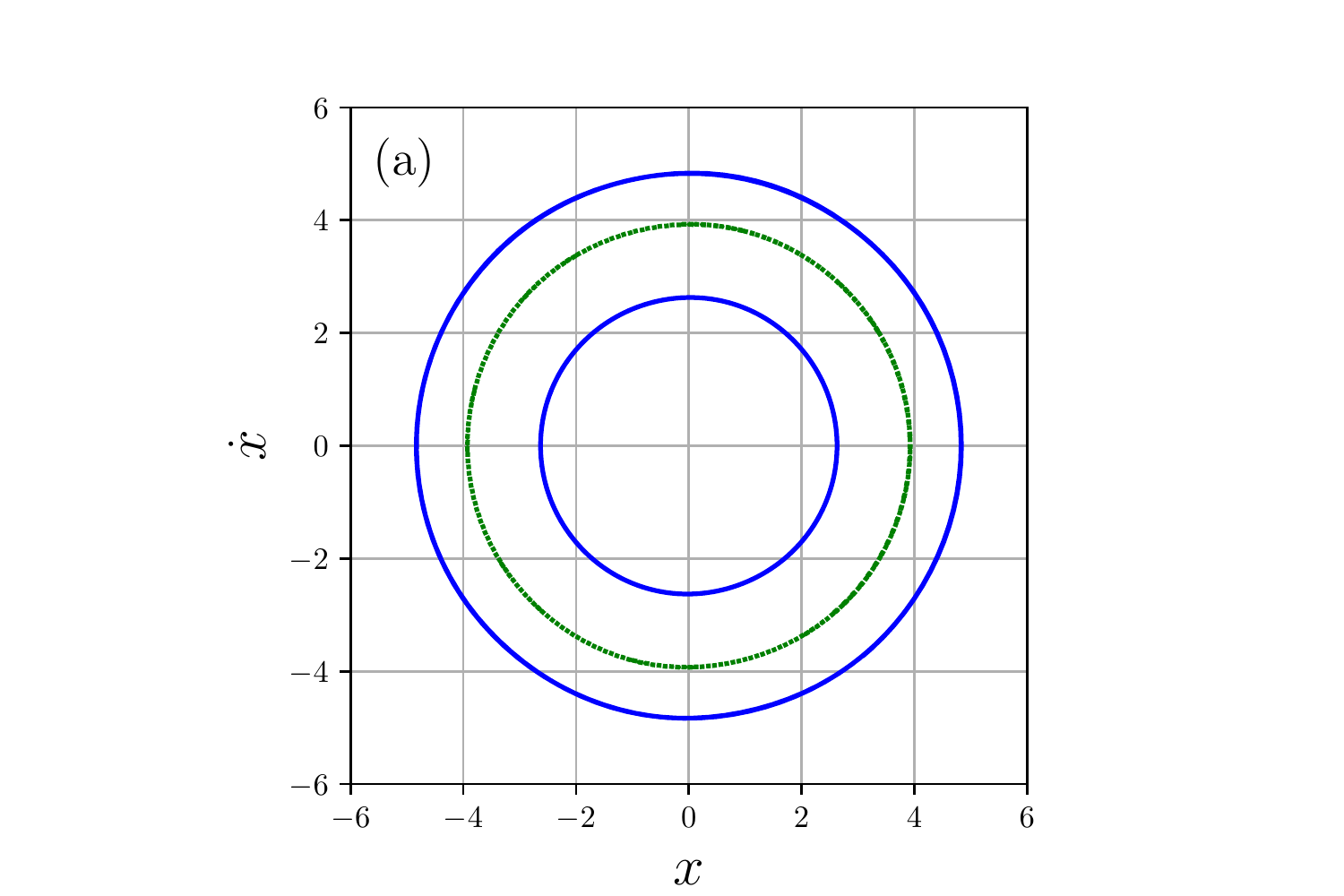}
		\includegraphics[width=1.15\columnwidth]{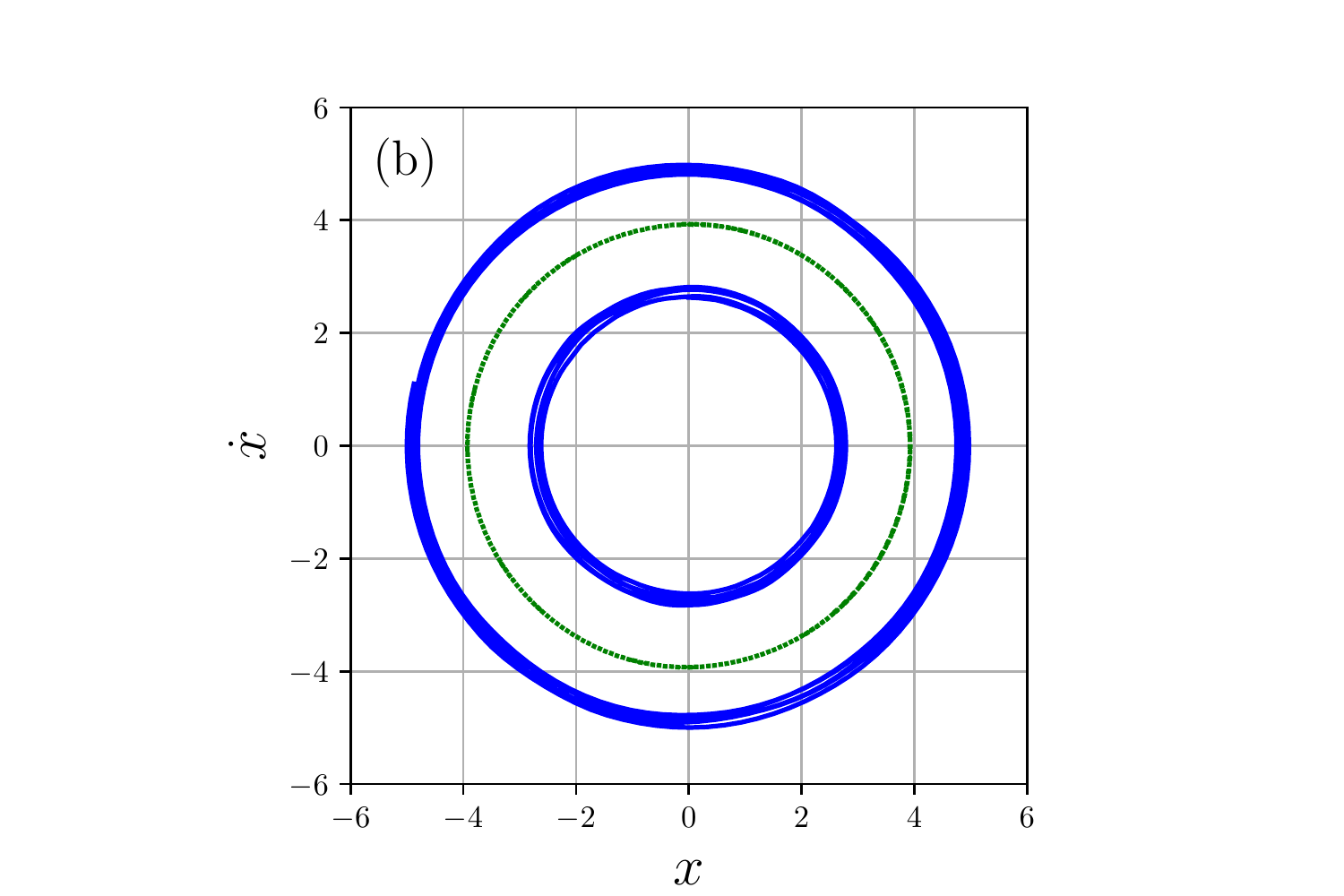}
\caption{ The phase space trajectories of (a) classical and (b) quantum Li\'{e}nard system with $f(x)=0.005*x^{6}-0.144*x^4+x^2-1$~\cite{saha_2020_systematic} are plotted for $\epsilon=0.01$. In both cases we have two stable (blue line) and one unstable (green dotted line) limit cycles.  Pure quantum fluctuations at $T=0$ may lead to noisy but stable oscillations even for the two stabe limit cycles (units arbitrary).} 
\label{Fig.IV} 
\end{figure}

 According to Refs.~\cite{blows_1984_number,giacomini_1997_number,libre_1998_limit,saha_2020_systematic} the system of Li\'{e}nard-Smith-Levinson form with $f(x)$, a polynomial of highest degree $2n$ where the coefficients, $a_n,a_{n-1}....a_1$ alternate in sign, can support at most $n$ number of limit cycles. The Eq.~\eqref{eom-gen-Leinard-osc} can be interpreted as quantum generalization of Li\'{e}nard oscillator with most general form of $f(x)$ which may have at most $n$ number of limit cycles. Phase portraits of a multiple limit cycle system in both classical and quantum scenario are plotted in Fig.~\ref{Fig.IV}. Classically, any noisy limit cycle at $T=0$ does not make any sense, but quantum mechanically, due to vacuum fluctuation one can have stable limit cycle oscillation even at absolute zero. This is a pure quantum phenomenon which does not have any classical analog. As an extension of Li\'{e}nard-Smith-Levinson form of oscillators, generalized Rayleigh family of oscillators~\cite{strutt_1877_theory,ghosh_2013_chemical,gaiko_2008_limit} can also be constructed from the above prescription (see Appendix).

\section{Conclusions}\label{Sec-V}

\par In this work we have presented a generalized microscopic quantum description of classical Li\'{e}nard system. Our approach is based on two essential elements. First, the interaction Hamiltonian consisting of suitable nonlinear coupling terms serves as a universal paradigm for description of several variants of Li\'{e}nard system with appropriate noise and nonlinear dissipation. Second, the harmonic oscillator coherent states facilitate the use of c-number description for reduced oscillator equation of motion.  The major findings of our work are as follows:

\par (i) A classical Li\'{e}nard system without any noise is purely a nonlinear dynamical oscillator which depending on the specificity of nonlinear dissipation admits of single or multiple limit cycles. As opposed to it, quantum noise appears in the dynamics as a consequence of system-reservoir framework of our microscopic model and carries the signature of pure quantum effects.

\par (ii) The effect of quantum noise makes the stable limit cycle fuzzy; however, it does not loose its stability in the sense that the limit cycle trajectory neither collapses to a steady state nor diverges. The noninterference of the two energy balance processes is characteristic of the internal noise of a thermodynamically closed system. Thus the detailed balance through nonlinear fluctuation-dissipation relation keeps the dynamical stability of the limit cycle preserved. In contrast, the external noise tends to destroy the limit cycle or, in general, the dynamical stability of the Li\'{e}nard system.

\par (iii) The Van der Pol and Rayleigh oscillators, though behave distinctly at the classical level, are identical at the microscopic level within a mean-field description. In fact a large number of Li\'{e}nard systems (both Van der Pol and Rayleigh families of cycles) differing in the form of their nonlinear damping can be constructed from the same microscopic Hamiltonian at the mean-field level. In all such  cases the number of limit cycles is determined by the highest power of the polynomial governing the nonlinear dissipation. 

\subsection*{ACKNOWLEDGEMENTS}

S. B. thanks DST INSPIRE, A. G. thanks Initiation Grant, IITK and SRG, SERB, India and D.S.R thanks DST, under Government of  India, for  a  J.  C.  Bose  National  Fellowship under Grant No. SB/S2/JCB-030/2015 for partial financial support.

\pagebreak
\onecolumngrid

\begin{center}
\textbf{APPENDIX}
\end{center}

\setcounter{equation}{0}
\renewcommand{\theequation}{A\arabic{equation}}
\subsection{Derivation of Eq.~\eqref{c-no-Langevin-eq-gen:n=1}}\label{Appendix-I}

Consider the operator Langevin equation~\eqref{op-langevin-eq:n=1} which is also the special case of generalized Langevin equation~\eqref{op-langevin-eq} for $n=1$. Taking average on both sides of the equation with the initial product separable coherent states of system and bath oscillators $\ket{\alpha}\ket{\mu_1}\ket{\mu_2}...\ket{\mu_N}$ one finds 
\begin{equation}\label{derv-c-no-Langevin: n=1, eq=1}
\dot{\alpha}=-\frac{\gamma_2}{m_1}(m_1 +m_0)\langle\alpha |\hat{A}^\dag \hat{A}^2 |\alpha \rangle +\frac{\gamma_2}{m_1}m_0 \langle\alpha |\hat{A}^\dag \hat{A}^2 |\alpha \rangle +f_2 \alpha^* ,
\end{equation}
where we have split $-\gamma_2 \langle\alpha|\hat{A}^\dag \hat{A}^2|\alpha\rangle$ into two terms by introducing two dimensionless numbers $m_0$ and $m_1$. $m_0$ and $m_1$ can assume positive or negative values depending on the specificity of the case and are independent of each other. This choice is determined by the characteristic polynomial of classical nonlinear damping. Rewriting the above equation ~\eqref{derv-c-no-Langevin: n=1, eq=1} we have
\begin{eqnarray}\label{derv-c-no-Langevin: n=1, eq=2}
\dot{\alpha} &=& -\frac{\gamma_2}{m_1}(m_1 +m_0)|\alpha |^2 \alpha  +\frac{\gamma_2}{m_1}m_0 \langle \alpha | \hat{A} (\hat{A}^\dag \hat{A} - 1) | \alpha \rangle + f_2 \alpha^* \nonumber \\
& \approx & -\frac{\gamma_2}{m_1}(m_1 +m_0)|\alpha |^2 \alpha   +\frac{\gamma_2}{m_1}m_0  \langle \alpha | \hat{A} | \alpha \rangle \langle \alpha |(\hat{A}^\dag \hat{A} -1) | \alpha \rangle + f_2 \alpha^* \nonumber \\
&=& -\frac{\gamma_2}{m_1}(m_1 +m_0)|\alpha |^2 \alpha  +\frac{\gamma_2}{m_1}m_0 (|\alpha |^2 -1)\alpha +f_2 \alpha^* \nonumber \\
&=& -\frac{\gamma_2}{m_1} (m_1 |\alpha|^2 + m_0)\alpha + f_2 \alpha^*. 
\end{eqnarray}
In deriving the Eq.~\eqref{derv-c-no-Langevin: n=1, eq=2} it has been assumed $\langle \hat{A} (\hat{A}^\dag \hat{A} - 1) \rangle \approx \langle \hat{A} \rangle \langle (\hat{A}^\dag \hat{A} -1) \rangle$, which is a valid approximation within the mean-field theory. This corresponds to neglect of higher order quantum correlation in the dynamics of nonlinear dissipation; quantum noise due to the heat-bath, however, completely unaffected.

\setcounter{equation}{0}
\renewcommand{\theequation}{B\arabic{equation}}\
\subsection{Derivation of Eq.~\eqref{c-no-Langevin-eq:n=2}}

Setting $n=2$ in Eq.~\eqref{op-langevin-eq} and averaging with $\ket{\alpha}\ket{\mu_1}\ket{\mu_2}...\ket{\mu_k}...\ket{\mu_N}$, we write down the c-number Langevin equation within the mean-field approximation as
\begin{eqnarray}\label{derv-c-no-Langevin-eq:n=2}
\dot{\alpha} &=&-\frac{\gamma_3}{m_2}(m_2+m_1) \langle \alpha |(\hat{A}^\dag)^2 \hat{A}^3 |\alpha\rangle+\frac{\gamma_3}{m_2}m_1 \langle \alpha |(\hat{A}^\dag)^2 \hat{A}^3 |\alpha\rangle + f_3 (\alpha^*)^2 \nonumber \\
&=&-\frac{\gamma_3}{m_2}(m_2+m_1) |\alpha|^4 \alpha +\frac{\gamma_3}{m_2}m_1\langle \alpha |\hat{A}^\dag (\hat{A}\hat{A}^\dag -1) \hat{A}^2|\alpha\rangle + f_3 (\alpha^*)^2 \nonumber\\
&\approx&-\frac{\gamma_3}{m_2}(m_2+m_1) |\alpha|^4 \alpha +\frac{\gamma_3}{m_2}m_1\langle \alpha |\hat{A}^\dag \hat{A}|\alpha\rangle \langle \alpha | \hat{A}^\dag \hat{A}^2 |\alpha \rangle -\frac{\gamma_3}{m_2}m_1 \langle\alpha | \hat{A}^\dag \hat{A}^2 | \alpha \rangle+ f_3 (\alpha^*)^2 \nonumber \\
&=&-\frac{\gamma_3}{m_2}(m_2+m_1) |\alpha|^4 \alpha +\frac{\gamma_3}{m_2}m_1 |\alpha |^4 \alpha-\frac{\gamma_3}{m_2}(m_1+m_0) \langle\alpha | \hat{A}^\dag \hat{A}^2 | \alpha \rangle +\frac{\gamma_3}{m_2}m_0 \langle \alpha |( \hat{A}\hat{A}^\dag \hat{A}-\hat{A})+ f_3 (\alpha^*)^2 \nonumber \\
&\approx&-\frac{\gamma_3}{m_2}m_2 |\alpha|^4 \alpha-\frac{\gamma_3}{m_2}(m_1 +m_0)|\alpha|^2 \alpha+\frac{\gamma_3}{m_2}m_0 \langle\alpha|\hat{A}|\alpha\rangle \langle \alpha|\hat{A}^\dag \hat{A}-1|\alpha\rangle+ f_3 (\alpha^*)^2 \nonumber \\
&=&-\frac{\gamma_3}{m_2}m_2 |\alpha|^4 \alpha-\frac{\gamma_3}{m_2}(m_1 +m_0)|\alpha|^2 \alpha+\frac{\gamma_3}{m_2}m_0(|\alpha|^2-1)\alpha+ f_3 (\alpha^*)^2 \nonumber \\
&=&-\frac{\gamma_3}{m_2}(m_2 |\alpha|^4 +m_1 |\alpha|^2 +m_0)\alpha+ f_3 (\alpha^*)^2 .
\end{eqnarray}
Here the mean-field approximation has been applied twice to obtain the Eq.~\eqref{derv-c-no-Langevin-eq:n=2} or Eq.~\eqref{c-no-Langevin-eq:n=2}. $m_2$, $m_1$ and $m_0$ are the integers to be chosen as per requirement of the form of the polynomial describing nonlinear dissipation.

\setcounter{equation}{0}
\renewcommand{\theequation}{C\arabic{equation}}\
\subsection{Derivation of Eq.~\eqref{c-no-Langevin-eq:gen}}

\par Following the same procedure as above, it may be anticipated that we can put $n=k$ in Eq.~\eqref{op-langevin-eq} and generate terms like $|\alpha|^{2k} \alpha$, $|\alpha|^{2k-2} \alpha .... $ upto $|\alpha|^2 \alpha$ using c-number formalism. The basis of this assumption lies on the previous two cases. Therefore we can presume the structure for $n=k$ within the purview of mean-field approximation as follows 
\begin{eqnarray}\label{c-no-Langevin-eq-C1}
\dot{\alpha}&=&-\gamma_{k+1}\langle \alpha | (\hat{A}^\dag) ^k \hat{A}^{k+1} | \alpha \rangle +f_{k+1}(\alpha^*)^k \nonumber\\
&=&-\frac{\gamma_{k+1}}{l_k}(l_k|\alpha|^{2k}+l_{k-1} |\alpha|^{2k-2}.....+l_1 |\alpha|^2+l_0)\alpha +f_{k+1}(\alpha^*)^k,
\end{eqnarray}
where the coefficients $l_k$, $l_{k-1}...$, $l_1$ and $l_0$ may or may not be interrelated.

\par Now having obtained the above equation for $n=k$, if we show that the same structure holds for $n=k+1$, we may claim that we prove our desired result Eq.~\eqref{c-no-Langevin-eq:gen}. For that we go back to Eq.~\eqref{op-langevin-eq} and  take $n=k+1$ and carry out the averaging with respect to coherent states as before
\begin{eqnarray}
\dot{\alpha}&=&-\gamma_{k+2} \langle \alpha | (\hat{A}^\dag)^{k+1} \hat{A}^{k+2}| \alpha \rangle +f_{k+2}(\alpha^*)^{k+1} \nonumber\\
&=&-\frac{\gamma_{k+2}}{m_{k+1}}(m_{k+1}-m)\langle \alpha | (\hat{A}^\dag)^{k+1} \hat{A}^{k+2}| \alpha \rangle -\frac{\gamma_{k+2}}{m_{k+1}}m \langle \alpha | (\hat{A}^\dag)^{k} (\hat{A}\hat{A}^\dag -1)\hat{A}^{k+1}| \alpha \rangle +f_{k+2}(\alpha^*)^{k+1}\nonumber \\
&\approx &-\frac{\gamma_{k+2}}{m_{k+1}}(m_{k+1}-m)|\alpha |^{2k+2} \alpha -\frac{\gamma_{k+2}}{m_{k+1}}m\langle \alpha | (\hat{A}^\dag)^{k} \hat{A} | \alpha \rangle\langle \alpha | \hat{A}^\dag\hat{A}^{k+1} | \alpha \rangle \nonumber \\ 
 &&+ \frac{\gamma_{k+2}}{m_{k+1}}m \langle \alpha | (\hat{A}^\dag)^{k} \hat{A}^{k+1} | \alpha \rangle +f_{k+2}(\alpha^*)^{k+1} \nonumber \\
 &=&-\frac{\gamma_{k+2}}{m_{k+1}}(m_{k+1}-m)|\alpha |^{2k+2} \alpha -\frac{\gamma_{k+2}}{m_{k+1}} m |\alpha|^{2k+2} \alpha \nonumber \\ 
 &&+\frac{\gamma_{k+2}}{m_{k+1}} \frac{m}{l_k}\{ l_k |\alpha|^{2k}+l_{k-1}|\alpha|^{2k-2}.....+l_1 |\alpha|^2+l_0 \} \alpha + f_{k+2} (\alpha^*)^{k+1}\label{c-no-Langevin-eq-gen-C2} \\
 &=&-\frac{\gamma_{k+2}}{m_{k+1}}(m_{k+1}-m)|\alpha |^{2k+2} \alpha -\frac{\gamma_{k+2}}{m_{k+1}}m  |\alpha|^{2k+2} \alpha \nonumber \\
 &&- \frac{\gamma_{k+2}}{m_{k+1}}(m_k |\alpha|^{2k}+m_{k-1} |\alpha|^{2k-2}...+ m_1 |\alpha|^2 +m_0 )\alpha+ f_{k+2} (\alpha^*)^{k+1} \label{c-no-Langevin-eq-gen-C3}\\ 
 &=& -\frac{\gamma_{k+2}}{m_{k+1}}(m_{k+1} |\alpha|^{2k+2}+m_k |\alpha |^{2k}... + m_1 |\alpha|+m_0)\alpha +f_{k+2} (\alpha^*)^{k+1}.\label{c-no-Langevin-eq-gen-C4}
\end{eqnarray}
Here we use Eq.~\eqref{c-no-Langevin-eq-C1} in Eq.~\eqref{c-no-Langevin-eq-gen-C2}. Further defining
\be
m_j=-\frac{l_{j}}{l_k}m, \quad \forall \quad  j=0,1,2 ... k. 
\ee
in Eq.~\eqref{c-no-Langevin-eq-gen-C3} we obtain Eq.~\eqref{c-no-Langevin-eq-gen-C4}. Thus we find the same structure for $n=k+1$ as for $n=k$. Therefore by the method of mathematical induction we have proved that the structure assumed in Eq.~\eqref{c-no-Langevin-eq:gen} is true for all $n \in {\mathbb{Z}}^+$. Similar technique can be followed to derive its complex conjugate equation.

\setcounter{equation}{0}
\renewcommand{\theequation}{D\arabic{equation}}\
\subsection{Rayleigh Family of Quantum Oscillators}

\begin{figure}[H]
	\centering
		\includegraphics[width=0.45\columnwidth]{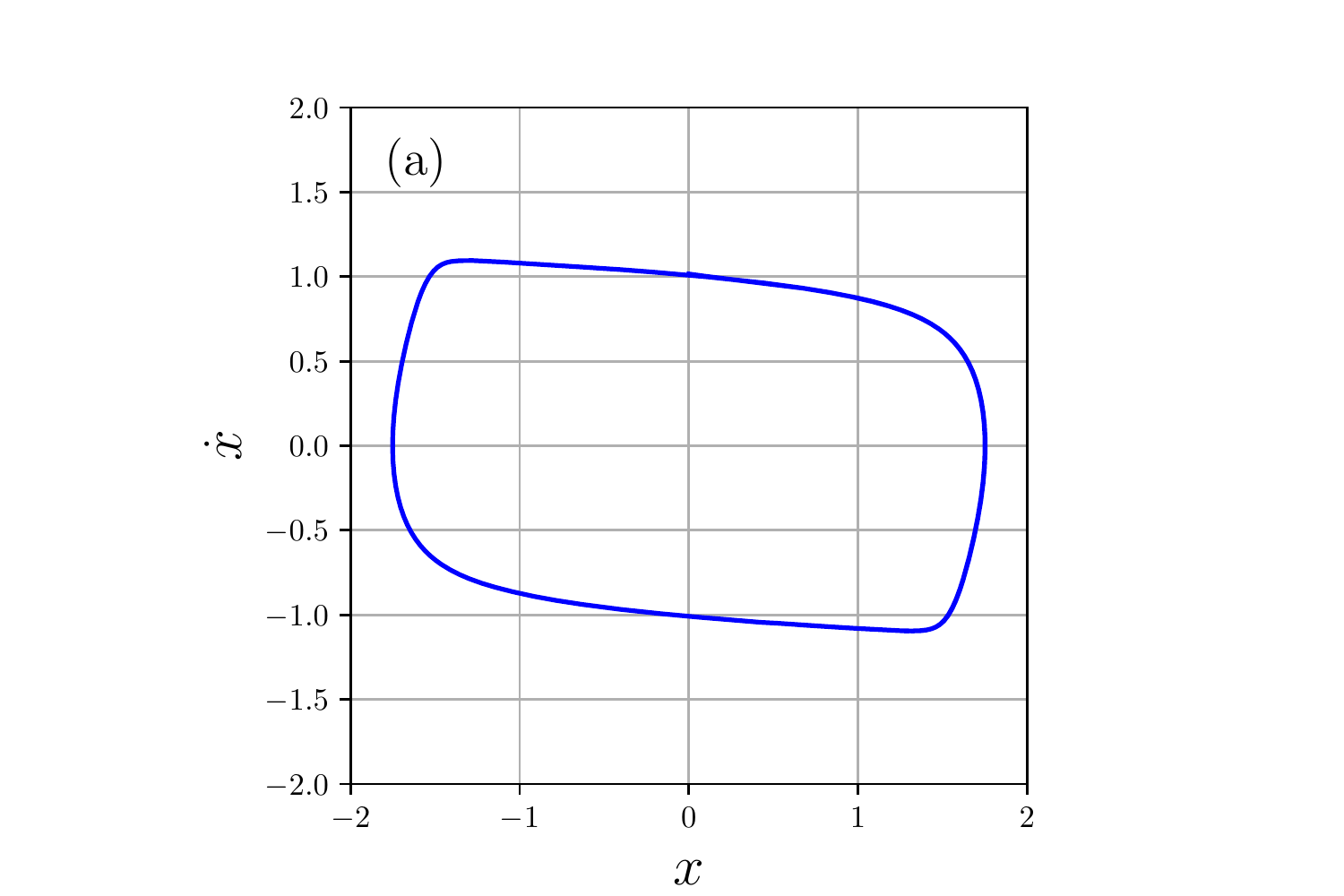}
		\includegraphics[width=0.45\columnwidth]{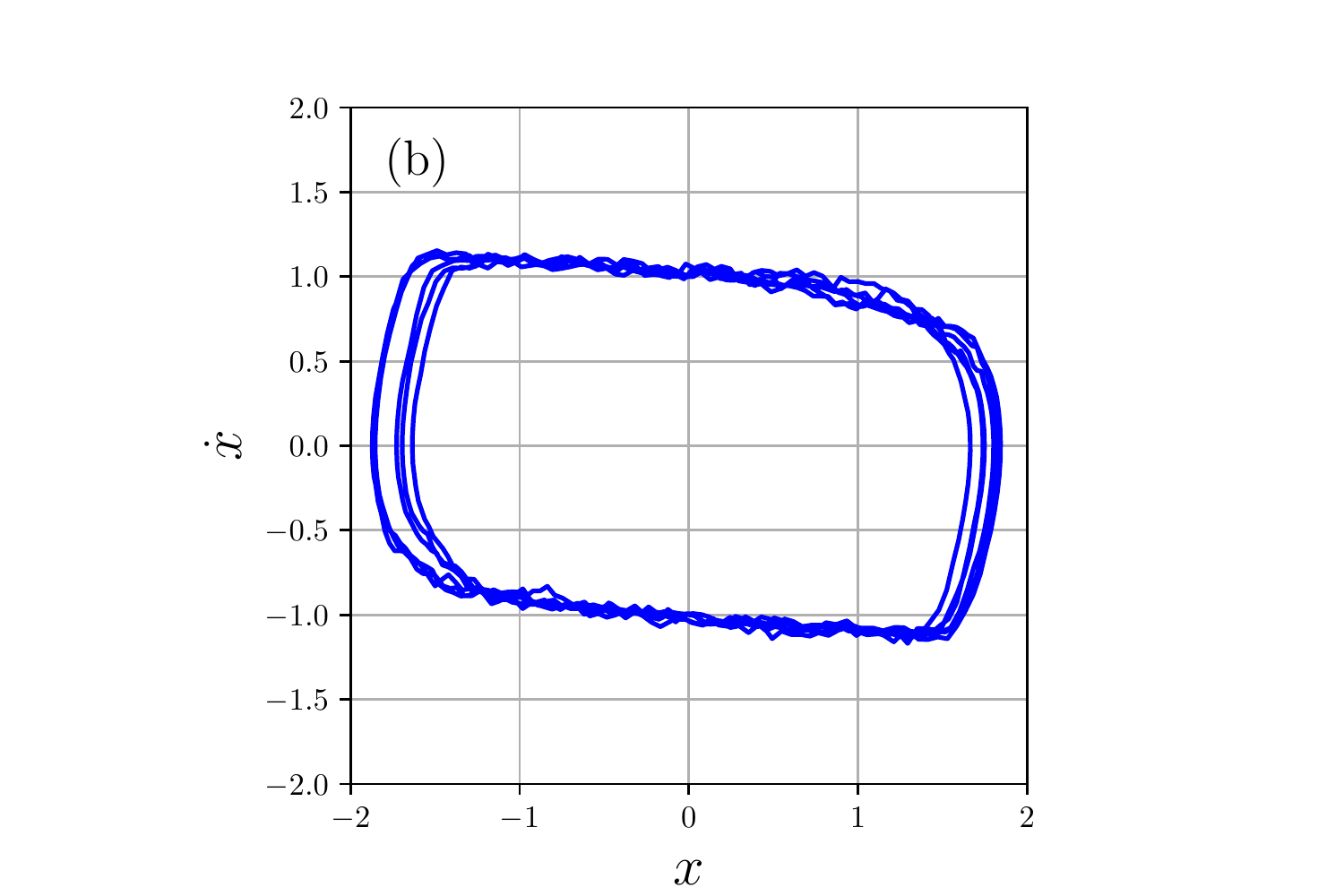}
\caption{ The phase space trajectories of (a) classical and (b) quantum version of a generalized Rayleigh oscillator, for polynomial ${{\dot{x}}^6+{\dot{x}}^4-{\dot{x}}^2-1}$ are plotted for $\epsilon=1.5$. We have put $KT=20$ for plotting the quantum case (units arbitrary).} \label{Fig.V} 
\end{figure}

From c-number Langevin  Eqs.\eqref{c-no-Langevin-eq:gen} and \eqref{c-no-Langevin-eq:gen-2} we construct the following form of the quantum equation with the help of substitution \eqref{substitution-rayleigh-osc}
\begin{equation}\label{family-Rayleigh-eq:gen}
\ddot{x}+\frac{\gamma_{n+1}}{m_n} \{ \beta_n {\dot{x}}^{2n}+\beta_{n-1}{\dot{x}}^{2n-2}.....+\beta_1{\dot{x}}^2+\beta_0 \}\dot{x} +\omega_0^2 x=\omega_0^{n-1} \eta_{n+1}.
\end{equation}
This Eq. \eqref{family-Rayleigh-eq:gen} can be considered as quantum analog of generalized Rayleigh family of oscillators~\cite{strutt_1877_theory,ghosh_2013_chemical}, where
\begin{equation}
\beta_j=\frac{m_j j}{^{2j}C_{j+1} (2j+1)},\quad  \forall \quad j=1,2,....n \quad \text{and} \quad\beta_0 =m_0.
\end{equation}
The family of oscillators could also support at most $n$ number of limit cycles, where $2n$ is the highest degree of the polynomial~\cite{gaiko_2008_limit}. The phase space trajectories of a representative example of such generalized Rayleigh oscillator is shown in both (a) classical and (b) quantum picture in Fig.~\eqref{Fig.V}.

\begin{thebibliography}{64}%
\makeatletter
\providecommand \@ifxundefined [1]{%
 \@ifx{#1\undefined}
}%
\providecommand \@ifnum [1]{%
 \ifnum #1\expandafter \@firstoftwo
 \else \expandafter \@secondoftwo
 \fi
}%
\providecommand \@ifx [1]{%
 \ifx #1\expandafter \@firstoftwo
 \else \expandafter \@secondoftwo
 \fi
}%
\providecommand \natexlab [1]{#1}%
\providecommand \enquote  [1]{``#1''}%
\providecommand \bibnamefont  [1]{#1}%
\providecommand \bibfnamefont [1]{#1}%
\providecommand \citenamefont [1]{#1}%
\providecommand \href@noop [0]{\@secondoftwo}%
\providecommand \href [0]{\begingroup \@sanitize@url \@href}%
\providecommand \@href[1]{\@@startlink{#1}\@@href}%
\providecommand \@@href[1]{\endgroup#1\@@endlink}%
\providecommand \@sanitize@url [0]{\catcode `\\12\catcode `\$12\catcode
  `\&12\catcode `\#12\catcode `\^12\catcode `\_12\catcode `\%12\relax}%
\providecommand \@@startlink[1]{}%
\providecommand \@@endlink[0]{}%
\providecommand \url  [0]{\begingroup\@sanitize@url \@url }%
\providecommand \@url [1]{\endgroup\@href {#1}{\urlprefix }}%
\providecommand \urlprefix  [0]{URL }%
\providecommand \Eprint [0]{\href }%
\providecommand \doibase [0]{http://dx.doi.org/}%
\providecommand \selectlanguage [0]{\@gobble}%
\providecommand \bibinfo  [0]{\@secondoftwo}%
\providecommand \bibfield  [0]{\@secondoftwo}%
\providecommand \translation [1]{[#1]}%
\providecommand \BibitemOpen [0]{}%
\providecommand \bibitemStop [0]{}%
\providecommand \bibitemNoStop [0]{.\EOS\space}%
\providecommand \EOS [0]{\spacefactor3000\relax}%
\providecommand \BibitemShut  [1]{\csname bibitem#1\endcsname}%
\let\auto@bib@innerbib\@empty
\bibitem [{\citenamefont {Lindenberg}\ and\ \citenamefont
  {West}(1990)}]{lindenberg_1990_nonequilibrium}%
  \BibitemOpen
  \bibfield  {author} {\bibinfo {author} {\bibfnamefont {K.}~\bibnamefont
  {Lindenberg}}\ and\ \bibinfo {author} {\bibfnamefont {B.}~\bibnamefont
  {West}},\ }\href {https://books.google.co.in/books?id=c6nvAAAAMAAJ} {\emph
  {\bibinfo {title} {The Nonequilibrium Statistical Mechanics of Open and
  Closed Systems}}}\ (\bibinfo  {publisher} {VCH Publishers},\ \bibinfo {year}
  {1990})\BibitemShut {NoStop}%
\bibitem [{\citenamefont {Weiss}(1999)}]{weiss_1999_quantum}%
  \BibitemOpen
  \bibfield  {author} {\bibinfo {author} {\bibfnamefont {U.}~\bibnamefont
  {Weiss}},\ }\href {https://books.google.co.in/books?id=kqZclKUZdq0C} {\emph
  {\bibinfo {title} {Quantum Dissipative Systems}}},\ Series in modern
  condensed matter physics\ (\bibinfo  {publisher} {World Scientific},\
  \bibinfo {year} {1999})\BibitemShut {NoStop}%
\bibitem [{\citenamefont {Louisell}(1973)}]{louisell_1973_quantum}%
  \BibitemOpen
  \bibfield  {author} {\bibinfo {author} {\bibfnamefont {W.}~\bibnamefont
  {Louisell}},\ }\href {https://books.google.co.in/books?id=NRlRAAAAMAAJ}
  {\emph {\bibinfo {title} {Quantum Statistical Properties of Radiation}}},\ A
  Wiley-Interscience publication\ (\bibinfo  {publisher} {Wiley},\ \bibinfo
  {year} {1973})\BibitemShut {NoStop}%
\bibitem [{\citenamefont {Carmichael}\ and\ \citenamefont
  {(Berlin).}(1999)}]{carmichael_1999_statistical_1}%
  \BibitemOpen
  \bibfield  {author} {\bibinfo {author} {\bibfnamefont {H.}~\bibnamefont
  {Carmichael}}\ and\ \bibinfo {author} {\bibfnamefont {S.-V.}\ \bibnamefont
  {(Berlin).}},\ }\href {https://books.google.co.in/books?id=ocgRgM-yJacC}
  {\emph {\bibinfo {title} {Statistical Methods in Quantum Optics 1: Master
  Equations and Fokker-Planck Equations}}},\ Physics and astronomy online
  library\ (\bibinfo  {publisher} {Springer},\ \bibinfo {year}
  {1999})\BibitemShut {NoStop}%
\bibitem [{\citenamefont {Zwanzig}(1973)}]{Zwanzig_1973_Nonlinear}%
  \BibitemOpen
  \bibfield  {author} {\bibinfo {author} {\bibfnamefont {R.}~\bibnamefont
  {Zwanzig}},\ }\href {\doibase 10.1007/BF01008729} {\bibfield  {journal}
  {\bibinfo  {journal} {Journal of Statistical Physics}\ }\textbf {\bibinfo
  {volume} {9}},\ \bibinfo {pages} {215} (\bibinfo {year} {1973})}\BibitemShut
  {NoStop}%
\bibitem [{\citenamefont {Caldeira}\ and\ \citenamefont
  {Leggett}(1981)}]{Calderia_1981_Influence}%
  \BibitemOpen
  \bibfield  {author} {\bibinfo {author} {\bibfnamefont {A.~O.}\ \bibnamefont
  {Caldeira}}\ and\ \bibinfo {author} {\bibfnamefont {A.~J.}\ \bibnamefont
  {Leggett}},\ }\href {\doibase 10.1103/PhysRevLett.46.211} {\bibfield
  {journal} {\bibinfo  {journal} {Phys. Rev. Lett.}\ }\textbf {\bibinfo
  {volume} {46}},\ \bibinfo {pages} {211} (\bibinfo {year} {1981})}\BibitemShut
  {NoStop}%
\bibitem [{\citenamefont {Caldeira}\ and\ \citenamefont
  {Leggett}(1983)}]{Calderia_1983_quantum}%
  \BibitemOpen
  \bibfield  {author} {\bibinfo {author} {\bibfnamefont {A.}~\bibnamefont
  {Caldeira}}\ and\ \bibinfo {author} {\bibfnamefont {A.}~\bibnamefont
  {Leggett}},\ }\href {\doibase https://doi.org/10.1016/0003-4916(83)90202-6}
  {\bibfield  {journal} {\bibinfo  {journal} {Annals of Physics}\ }\textbf
  {\bibinfo {volume} {149}},\ \bibinfo {pages} {374 } (\bibinfo {year}
  {1983})}\BibitemShut {NoStop}%
\bibitem [{\citenamefont {Leggett}\ \emph {et~al.}(1987)\citenamefont
  {Leggett}, \citenamefont {Chakravarty}, \citenamefont {Dorsey}, \citenamefont
  {Fisher}, \citenamefont {Garg},\ and\ \citenamefont
  {Zwerger}}]{Calderia_1987_dynamics}%
  \BibitemOpen
  \bibfield  {author} {\bibinfo {author} {\bibfnamefont {A.~J.}\ \bibnamefont
  {Leggett}}, \bibinfo {author} {\bibfnamefont {S.}~\bibnamefont
  {Chakravarty}}, \bibinfo {author} {\bibfnamefont {A.~T.}\ \bibnamefont
  {Dorsey}}, \bibinfo {author} {\bibfnamefont {M.~P.~A.}\ \bibnamefont
  {Fisher}}, \bibinfo {author} {\bibfnamefont {A.}~\bibnamefont {Garg}}, \ and\
  \bibinfo {author} {\bibfnamefont {W.}~\bibnamefont {Zwerger}},\ }\href
  {\doibase 10.1103/RevModPhys.59.1} {\bibfield  {journal} {\bibinfo  {journal}
  {Rev. Mod. Phys.}\ }\textbf {\bibinfo {volume} {59}},\ \bibinfo {pages} {1}
  (\bibinfo {year} {1987})}\BibitemShut {NoStop}%
\bibitem [{\citenamefont {Nitzan}(2013)}]{nitzan_2013_chemical}%
  \BibitemOpen
  \bibfield  {author} {\bibinfo {author} {\bibfnamefont {A.}~\bibnamefont
  {Nitzan}},\ }\href {https://books.google.co.in/books?id=jrmznAEACAAJ} {\emph
  {\bibinfo {title} {Chemical Dynamics in Condensed Phases: Relaxation,
  Transfer, and Reactions in Condensed Molecular Systems}}},\ Oxford Graduate
  Texts\ (\bibinfo  {publisher} {OUP Oxford},\ \bibinfo {year}
  {2013})\BibitemShut {NoStop}%
\bibitem [{\citenamefont {Agarwal}(2013)}]{agarwalbookquantumoptics}%
  \BibitemOpen
  \bibfield  {author} {\bibinfo {author} {\bibfnamefont {G.~S.}\ \bibnamefont
  {Agarwal}},\ }\href@noop {} {\emph {\bibinfo {title} {Quantum Optics}}}\
  (\bibinfo  {publisher} {Cambridge University Press},\ \bibinfo {address}
  {Cambridge},\ \bibinfo {year} {2013})\BibitemShut {NoStop}%
\bibitem [{\citenamefont {Munowitz}(1988)}]{munowitz_1988_coherence}%
  \BibitemOpen
  \bibfield  {author} {\bibinfo {author} {\bibfnamefont {M.}~\bibnamefont
  {Munowitz}},\ }\href {https://books.google.co.in/books?id=h9LvAAAAMAAJ}
  {\emph {\bibinfo {title} {Coherence and NMR}}}\ (\bibinfo  {publisher}
  {Wiley},\ \bibinfo {year} {1988})\BibitemShut {NoStop}%
\bibitem [{\citenamefont {H\"anggi}\ \emph {et~al.}(1990)\citenamefont
  {H\"anggi}, \citenamefont {Talkner},\ and\ \citenamefont
  {Borkovec}}]{hanggi_1990_reaction}%
  \BibitemOpen
  \bibfield  {author} {\bibinfo {author} {\bibfnamefont {P.}~\bibnamefont
  {H\"anggi}}, \bibinfo {author} {\bibfnamefont {P.}~\bibnamefont {Talkner}}, \
  and\ \bibinfo {author} {\bibfnamefont {M.}~\bibnamefont {Borkovec}},\ }\href
  {\doibase 10.1103/RevModPhys.62.251} {\bibfield  {journal} {\bibinfo
  {journal} {Rev. Mod. Phys.}\ }\textbf {\bibinfo {volume} {62}},\ \bibinfo
  {pages} {251} (\bibinfo {year} {1990})}\BibitemShut {NoStop}%
\bibitem [{\citenamefont {Hern\'andez}\ and\ \citenamefont
  {Dorso}(1984)}]{hernandez_1984_quantal}%
  \BibitemOpen
  \bibfield  {author} {\bibinfo {author} {\bibfnamefont {E.~S.}\ \bibnamefont
  {Hern\'andez}}\ and\ \bibinfo {author} {\bibfnamefont {C.~O.}\ \bibnamefont
  {Dorso}},\ }\href {\doibase 10.1103/PhysRevC.29.1510} {\bibfield  {journal}
  {\bibinfo  {journal} {Phys. Rev. C}\ }\textbf {\bibinfo {volume} {29}},\
  \bibinfo {pages} {1510} (\bibinfo {year} {1984})}\BibitemShut {NoStop}%
\bibitem [{\citenamefont {Vlad\'ar}\ \emph {et~al.}(1986)\citenamefont
  {Vlad\'ar}, \citenamefont {Zim\'anyi},\ and\ \citenamefont
  {Zawadowski}}]{vladar_1986_theory}%
  \BibitemOpen
  \bibfield  {author} {\bibinfo {author} {\bibfnamefont {K.}~\bibnamefont
  {Vlad\'ar}}, \bibinfo {author} {\bibfnamefont {G.~T.}\ \bibnamefont
  {Zim\'anyi}}, \ and\ \bibinfo {author} {\bibfnamefont {A.}~\bibnamefont
  {Zawadowski}},\ }\href {\doibase 10.1103/PhysRevLett.56.286} {\bibfield
  {journal} {\bibinfo  {journal} {Phys. Rev. Lett.}\ }\textbf {\bibinfo
  {volume} {56}},\ \bibinfo {pages} {286} (\bibinfo {year} {1986})}\BibitemShut
  {NoStop}%
\bibitem [{\citenamefont {Caldeira}\ \emph {et~al.}(1993)\citenamefont
  {Caldeira}, \citenamefont {Castro~Neto},\ and\ \citenamefont {Oliveira~de
  Carvalho}}]{calderia_1993_dissipative}%
  \BibitemOpen
  \bibfield  {author} {\bibinfo {author} {\bibfnamefont {A.~O.}\ \bibnamefont
  {Caldeira}}, \bibinfo {author} {\bibfnamefont {A.~H.}\ \bibnamefont
  {Castro~Neto}}, \ and\ \bibinfo {author} {\bibfnamefont {T.}~\bibnamefont
  {Oliveira~de Carvalho}},\ }\href {\doibase 10.1103/PhysRevB.48.13974}
  {\bibfield  {journal} {\bibinfo  {journal} {Phys. Rev. B}\ }\textbf {\bibinfo
  {volume} {48}},\ \bibinfo {pages} {13974} (\bibinfo {year}
  {1993})}\BibitemShut {NoStop}%
\bibitem [{\citenamefont {Guinea}\ \emph {et~al.}(1985)\citenamefont {Guinea},
  \citenamefont {Hakim},\ and\ \citenamefont
  {Muramatsu}}]{guinea_1985_bosonization}%
  \BibitemOpen
  \bibfield  {author} {\bibinfo {author} {\bibfnamefont {F.}~\bibnamefont
  {Guinea}}, \bibinfo {author} {\bibfnamefont {V.}~\bibnamefont {Hakim}}, \
  and\ \bibinfo {author} {\bibfnamefont {A.}~\bibnamefont {Muramatsu}},\ }\href
  {\doibase 10.1103/PhysRevB.32.4410} {\bibfield  {journal} {\bibinfo
  {journal} {Phys. Rev. B}\ }\textbf {\bibinfo {volume} {32}},\ \bibinfo
  {pages} {4410} (\bibinfo {year} {1985})}\BibitemShut {NoStop}%
\bibitem [{\citenamefont {Ferrer}\ and\ \citenamefont
  {Smith}(2007)}]{ferrer_2007_dynamical}%
  \BibitemOpen
  \bibfield  {author} {\bibinfo {author} {\bibfnamefont {A.~V.}\ \bibnamefont
  {Ferrer}}\ and\ \bibinfo {author} {\bibfnamefont {C.~M.}\ \bibnamefont
  {Smith}},\ }\href {\doibase 10.1103/PhysRevB.76.214303} {\bibfield  {journal}
  {\bibinfo  {journal} {Phys. Rev. B}\ }\textbf {\bibinfo {volume} {76}},\
  \bibinfo {pages} {214303} (\bibinfo {year} {2007})}\BibitemShut {NoStop}%
\bibitem [{\citenamefont {Villares~Ferrer}\ \emph {et~al.}(2006)\citenamefont
  {Villares~Ferrer}, \citenamefont {Caldeira},\ and\ \citenamefont
  {Smith}}]{villares_2006_optical}%
  \BibitemOpen
  \bibfield  {author} {\bibinfo {author} {\bibfnamefont {A.}~\bibnamefont
  {Villares~Ferrer}}, \bibinfo {author} {\bibfnamefont {A.~O.}\ \bibnamefont
  {Caldeira}}, \ and\ \bibinfo {author} {\bibfnamefont {C.~M.}\ \bibnamefont
  {Smith}},\ }\href {\doibase 10.1103/PhysRevB.74.184304} {\bibfield  {journal}
  {\bibinfo  {journal} {Phys. Rev. B}\ }\textbf {\bibinfo {volume} {74}},\
  \bibinfo {pages} {184304} (\bibinfo {year} {2006})}\BibitemShut {NoStop}%
\bibitem [{\citenamefont {Strogatz}(2014)}]{strogatz_2014_nonlinear}%
  \BibitemOpen
  \bibfield  {author} {\bibinfo {author} {\bibfnamefont {S.}~\bibnamefont
  {Strogatz}},\ }\href {https://books.google.co.in/books?id=JDQGAwAAQBAJ}
  {\emph {\bibinfo {title} {Nonlinear Dynamics and Chaos: With Applications to
  Physics, Biology, Chemistry, and Engineering}}},\ Studies in Nonlinearity\
  (\bibinfo  {publisher} {Avalon Publishing},\ \bibinfo {year}
  {2014})\BibitemShut {NoStop}%
\bibitem [{\citenamefont {der Pol}(1920)}]{b_van_1920}%
  \BibitemOpen
  \bibfield  {author} {\bibinfo {author} {\bibfnamefont {B.}~\bibnamefont {der
  Pol}},\ }\href@noop {} {\bibfield  {journal} {\bibinfo  {journal} {Radio
  Review}\ }\textbf {\bibinfo {volume} {1}},\ \bibinfo {pages} {701} (\bibinfo
  {year} {1920})}\BibitemShut {NoStop}%
\bibitem [{\citenamefont {van~der Pol Jun.~D.Sc.}(1922)}]{b_van_1922}%
  \BibitemOpen
  \bibfield  {author} {\bibinfo {author} {\bibfnamefont {B.}~\bibnamefont
  {van~der Pol Jun.~D.Sc.}},\ }\href {\doibase 10.1080/14786442208633932}
  {\bibfield  {journal} {\bibinfo  {journal} {The London, Edinburgh, and Dublin
  Philosophical Magazine and Journal of Science}\ }\textbf {\bibinfo {volume}
  {43}},\ \bibinfo {pages} {700} (\bibinfo {year} {1922})}\BibitemShut
  {NoStop}%
\bibitem [{\citenamefont {Strutt (Lord~Rayleigh)}(1877)}]{strutt_1877_theory}%
  \BibitemOpen
  \bibfield  {author} {\bibinfo {author} {\bibfnamefont {J.}~\bibnamefont
  {Strutt (Lord~Rayleigh)}},\ }\href
  {https://books.google.co.in/books?id=fj0DAAAAQAAJ} {\emph {\bibinfo {title}
  {The theory of sound}}},\ \bibinfo {series} {The theory of sound}\ No.\
  \bibinfo {number} {v. 1}\ (\bibinfo  {publisher} {Macmillan and Company},\
  \bibinfo {year} {1877})\BibitemShut {NoStop}%
\bibitem [{\citenamefont {Levinson}\ and\ \citenamefont
  {Smith}(1942)}]{levinson1942general}%
  \BibitemOpen
  \bibfield  {author} {\bibinfo {author} {\bibfnamefont {N.}~\bibnamefont
  {Levinson}}\ and\ \bibinfo {author} {\bibfnamefont {O.~K.}\ \bibnamefont
  {Smith}},\ }\href {\doibase 10.1215/S0012-7094-42-00928-1} {\bibfield
  {journal} {\bibinfo  {journal} {Duke Math. J.}\ }\textbf {\bibinfo {volume}
  {9}},\ \bibinfo {pages} {382} (\bibinfo {year} {1942})}\BibitemShut {NoStop}%
\bibitem [{\citenamefont {Saha}\ \emph {et~al.}(2019)\citenamefont {Saha},
  \citenamefont {Gangopadhyay},\ and\ \citenamefont
  {Ray}}]{saha_2019_reduction}%
  \BibitemOpen
  \bibfield  {author} {\bibinfo {author} {\bibfnamefont {S.}~\bibnamefont
  {Saha}}, \bibinfo {author} {\bibfnamefont {G.}~\bibnamefont {Gangopadhyay}},
  \ and\ \bibinfo {author} {\bibfnamefont {D.~S.}\ \bibnamefont {Ray}},\ }\href
  {\doibase 10.1007/s40819-019-0628-9} {\bibfield  {journal} {\bibinfo
  {journal} {International Journal of Applied and Computational Mathematics}\
  }\textbf {\bibinfo {volume} {5}},\ \bibinfo {pages} {46} (\bibinfo {year}
  {2019})}\BibitemShut {NoStop}%
\bibitem [{\citenamefont {Saha}\ \emph {et~al.}(2020)\citenamefont {Saha},
  \citenamefont {Gangopadhyay},\ and\ \citenamefont {{Shankar
  Ray}}}]{saha_2020_systematic}%
  \BibitemOpen
  \bibfield  {author} {\bibinfo {author} {\bibfnamefont {S.}~\bibnamefont
  {Saha}}, \bibinfo {author} {\bibfnamefont {G.}~\bibnamefont {Gangopadhyay}},
  \ and\ \bibinfo {author} {\bibfnamefont {D.}~\bibnamefont {{Shankar Ray}}},\
  }\href {\doibase https://doi.org/10.1016/j.cnsns.2020.105234} {\bibfield
  {journal} {\bibinfo  {journal} {Communications in Nonlinear Science and
  Numerical Simulation}\ }\textbf {\bibinfo {volume} {85}},\ \bibinfo {pages}
  {105234} (\bibinfo {year} {2020})}\BibitemShut {NoStop}%
\bibitem [{\citenamefont {Landau}\ and\ \citenamefont
  {Lifshitz}(1980)}]{landaubook}%
  \BibitemOpen
  \bibfield  {author} {\bibinfo {author} {\bibfnamefont {L.~D.}\ \bibnamefont
  {Landau}}\ and\ \bibinfo {author} {\bibfnamefont {E.~M.}\ \bibnamefont
  {Lifshitz}},\ }\href@noop {} {\emph {\bibinfo {title} {Statistical Physics
  Part 1}}},\ \bibinfo {edition} {3rd}\ ed.\ (\bibinfo  {publisher} {Pergamon
  Press Ltd.},\ \bibinfo {address} {Oxford},\ \bibinfo {year}
  {1980})\BibitemShut {NoStop}%
\bibitem [{\citenamefont {Callen}(1985)}]{callenbook}%
  \BibitemOpen
  \bibfield  {author} {\bibinfo {author} {\bibfnamefont {H.~B.}\ \bibnamefont
  {Callen}},\ }\href@noop {} {\emph {\bibinfo {title} {{Thermodynamics and an
  Introduction to Thermostatistics}}}},\ \bibinfo {edition} {2nd}\ ed.\
  (\bibinfo  {publisher} {John Wiley \& Sons, Inc.},\ \bibinfo {address} {New
  York},\ \bibinfo {year} {1985})\BibitemShut {NoStop}%
\bibitem [{\citenamefont {Owen}\ \emph {et~al.}(2018)\citenamefont {Owen},
  \citenamefont {Jin}, \citenamefont {Rossini}, \citenamefont {Fazio},\ and\
  \citenamefont {Hartmann}}]{Owen_2018_quantum}%
  \BibitemOpen
  \bibfield  {author} {\bibinfo {author} {\bibfnamefont {E.~T.}\ \bibnamefont
  {Owen}}, \bibinfo {author} {\bibfnamefont {J.}~\bibnamefont {Jin}}, \bibinfo
  {author} {\bibfnamefont {D.}~\bibnamefont {Rossini}}, \bibinfo {author}
  {\bibfnamefont {R.}~\bibnamefont {Fazio}}, \ and\ \bibinfo {author}
  {\bibfnamefont {M.~J.}\ \bibnamefont {Hartmann}},\ }\href {\doibase
  10.1088/1367-2630/aab7d3} {\bibfield  {journal} {\bibinfo  {journal} {New
  Journal of Physics}\ }\textbf {\bibinfo {volume} {20}},\ \bibinfo {pages}
  {045004} (\bibinfo {year} {2018})}\BibitemShut {NoStop}%
\bibitem [{\citenamefont {Navarrete-Benlloch}\ \emph
  {et~al.}(2017)\citenamefont {Navarrete-Benlloch}, \citenamefont {Weiss},
  \citenamefont {Walter},\ and\ \citenamefont
  {de~Valc\'arcel}}]{navarrete_2017_general}%
  \BibitemOpen
  \bibfield  {author} {\bibinfo {author} {\bibfnamefont {C.}~\bibnamefont
  {Navarrete-Benlloch}}, \bibinfo {author} {\bibfnamefont {T.}~\bibnamefont
  {Weiss}}, \bibinfo {author} {\bibfnamefont {S.}~\bibnamefont {Walter}}, \
  and\ \bibinfo {author} {\bibfnamefont {G.~J.}\ \bibnamefont
  {de~Valc\'arcel}},\ }\href {\doibase 10.1103/PhysRevLett.119.133601}
  {\bibfield  {journal} {\bibinfo  {journal} {Phys. Rev. Lett.}\ }\textbf
  {\bibinfo {volume} {119}},\ \bibinfo {pages} {133601} (\bibinfo {year}
  {2017})}\BibitemShut {NoStop}%
\bibitem [{\citenamefont {Hillery}\ \emph {et~al.}(1984)\citenamefont
  {Hillery}, \citenamefont {O'Connell}, \citenamefont {Scully},\ and\
  \citenamefont {Wigner}}]{hilley_1984_distribution}%
  \BibitemOpen
  \bibfield  {author} {\bibinfo {author} {\bibfnamefont {M.}~\bibnamefont
  {Hillery}}, \bibinfo {author} {\bibfnamefont {R.}~\bibnamefont {O'Connell}},
  \bibinfo {author} {\bibfnamefont {M.}~\bibnamefont {Scully}}, \ and\ \bibinfo
  {author} {\bibfnamefont {E.}~\bibnamefont {Wigner}},\ }\href {\doibase
  https://doi.org/10.1016/0370-1573(84)90160-1} {\bibfield  {journal} {\bibinfo
   {journal} {Physics Reports}\ }\textbf {\bibinfo {volume} {106}},\ \bibinfo
  {pages} {121 } (\bibinfo {year} {1984})}\BibitemShut {NoStop}%
\bibitem [{\citenamefont {Barik}\ \emph {et~al.}(2005)\citenamefont {Barik},
  \citenamefont {Banerjee},\ and\ \citenamefont {Ray}}]{barik_2005_quantum}%
  \BibitemOpen
  \bibfield  {author} {\bibinfo {author} {\bibfnamefont {D.}~\bibnamefont
  {Barik}}, \bibinfo {author} {\bibfnamefont {D.}~\bibnamefont {Banerjee}}, \
  and\ \bibinfo {author} {\bibfnamefont {D.}~\bibnamefont {Ray}},\ }\href
  {https://books.google.co.in/books?id=I04oAQAACAAJ} {\emph {\bibinfo {title}
  {Quantum Brownian Motion in C-numbers: Theory and Applications}}}\ (\bibinfo
  {publisher} {Nova Science Publishers, Incorporated},\ \bibinfo {year}
  {2005})\BibitemShut {NoStop}%
\bibitem [{\citenamefont {Ghosh}\ \emph {et~al.}(2012)\citenamefont {Ghosh},
  \citenamefont {Sinha},\ and\ \citenamefont {Ray}}]{ghosh_2012_canonical}%
  \BibitemOpen
  \bibfield  {author} {\bibinfo {author} {\bibfnamefont {A.}~\bibnamefont
  {Ghosh}}, \bibinfo {author} {\bibfnamefont {S.~S.}\ \bibnamefont {Sinha}}, \
  and\ \bibinfo {author} {\bibfnamefont {D.~S.}\ \bibnamefont {Ray}},\ }\href
  {\doibase 10.1103/PhysRevE.86.011122} {\bibfield  {journal} {\bibinfo
  {journal} {Phys. Rev. E}\ }\textbf {\bibinfo {volume} {86}},\ \bibinfo
  {pages} {011122} (\bibinfo {year} {2012})}\BibitemShut {NoStop}%
\bibitem [{\citenamefont {Banerjee}\ \emph {et~al.}(2002)\citenamefont
  {Banerjee}, \citenamefont {Bag}, \citenamefont {Banik},\ and\ \citenamefont
  {Ray}}]{banerjee_2002_approach}%
  \BibitemOpen
  \bibfield  {author} {\bibinfo {author} {\bibfnamefont {D.}~\bibnamefont
  {Banerjee}}, \bibinfo {author} {\bibfnamefont {B.~C.}\ \bibnamefont {Bag}},
  \bibinfo {author} {\bibfnamefont {S.~K.}\ \bibnamefont {Banik}}, \ and\
  \bibinfo {author} {\bibfnamefont {D.~S.}\ \bibnamefont {Ray}},\ }\href
  {\doibase 10.1103/PhysRevE.65.021109} {\bibfield  {journal} {\bibinfo
  {journal} {Phys. Rev. E}\ }\textbf {\bibinfo {volume} {65}},\ \bibinfo
  {pages} {021109} (\bibinfo {year} {2002})}\BibitemShut {NoStop}%
\bibitem [{\citenamefont {Sinha}\ \emph
  {et~al.}(2011{\natexlab{a}})\citenamefont {Sinha}, \citenamefont {Ghosh},\
  and\ \citenamefont {Ray}}]{sinha2011quantum}%
  \BibitemOpen
  \bibfield  {author} {\bibinfo {author} {\bibfnamefont {S.~S.}\ \bibnamefont
  {Sinha}}, \bibinfo {author} {\bibfnamefont {A.}~\bibnamefont {Ghosh}}, \ and\
  \bibinfo {author} {\bibfnamefont {D.~S.}\ \bibnamefont {Ray}},\ }\href
  {\doibase 10.1103/PhysRevE.84.031118} {\bibfield  {journal} {\bibinfo
  {journal} {Phys. Rev. E}\ }\textbf {\bibinfo {volume} {84}},\ \bibinfo
  {pages} {031118} (\bibinfo {year} {2011}{\natexlab{a}})}\BibitemShut
  {NoStop}%
\bibitem [{\citenamefont {Sinha}\ \emph
  {et~al.}(2011{\natexlab{b}})\citenamefont {Sinha}, \citenamefont {Ghosh},\
  and\ \citenamefont {Ray}}]{sinha_2011_decay}%
  \BibitemOpen
  \bibfield  {author} {\bibinfo {author} {\bibfnamefont {S.~S.}\ \bibnamefont
  {Sinha}}, \bibinfo {author} {\bibfnamefont {A.}~\bibnamefont {Ghosh}}, \ and\
  \bibinfo {author} {\bibfnamefont {D.~S.}\ \bibnamefont {Ray}},\ }\href
  {\doibase 10.1103/PhysRevE.84.041113} {\bibfield  {journal} {\bibinfo
  {journal} {Phys. Rev. E}\ }\textbf {\bibinfo {volume} {84}},\ \bibinfo
  {pages} {041113} (\bibinfo {year} {2011}{\natexlab{b}})}\BibitemShut
  {NoStop}%
\bibitem [{\citenamefont {Ghosh}\ \emph {et~al.}(2011)\citenamefont {Ghosh},
  \citenamefont {Sinha},\ and\ \citenamefont {Ray}}]{ghosh2011dissipation}%
  \BibitemOpen
  \bibfield  {author} {\bibinfo {author} {\bibfnamefont {A.}~\bibnamefont
  {Ghosh}}, \bibinfo {author} {\bibfnamefont {S.~S.}\ \bibnamefont {Sinha}}, \
  and\ \bibinfo {author} {\bibfnamefont {D.~S.}\ \bibnamefont {Ray}},\ }\href
  {\doibase 10.1103/PhysRevE.83.061154} {\bibfield  {journal} {\bibinfo
  {journal} {Phys. Rev. E}\ }\textbf {\bibinfo {volume} {83}},\ \bibinfo
  {pages} {061154} (\bibinfo {year} {2011})}\BibitemShut {NoStop}%
\bibitem [{\citenamefont {Sinha}\ \emph {et~al.}(2013)\citenamefont {Sinha},
  \citenamefont {Ghosh}, \citenamefont {Bag},\ and\ \citenamefont
  {Ray}}]{sinha2013brownian}%
  \BibitemOpen
  \bibfield  {author} {\bibinfo {author} {\bibfnamefont {S.~S.}\ \bibnamefont
  {Sinha}}, \bibinfo {author} {\bibfnamefont {A.}~\bibnamefont {Ghosh}},
  \bibinfo {author} {\bibfnamefont {B.~C.}\ \bibnamefont {Bag}}, \ and\
  \bibinfo {author} {\bibfnamefont {D.~S.}\ \bibnamefont {Ray}},\ }\enquote
  {\bibinfo {title} {Quantum brownian motion in a spin bath},}\ in\ \href
  {https://books.google.co.in/books?id=s5TFAQAACAAJ} {\emph {\bibinfo
  {booktitle} {Concepts and Methods in Modern Theoretical Chemistry, Volume
  1}}},\ \bibinfo {editor} {edited by\ \bibinfo {editor} {\bibfnamefont
  {S.~K.}\ \bibnamefont {Ghosh}}\ and\ \bibinfo {editor} {\bibfnamefont
  {P.~K.}\ \bibnamefont {Chattaraj}}}\ (\bibinfo  {publisher} {Taylor \&
  Francis},\ \bibinfo {address} {Cham},\ \bibinfo {year} {2013})\ pp.\ \bibinfo
  {pages} {183--204}\BibitemShut {NoStop}%
\bibitem [{\citenamefont {Goldbeter}\ and\ \citenamefont
  {Berridge}(1997)}]{goldbeter_1997_biochemical}%
  \BibitemOpen
  \bibfield  {author} {\bibinfo {author} {\bibfnamefont {A.}~\bibnamefont
  {Goldbeter}}\ and\ \bibinfo {author} {\bibfnamefont {M.}~\bibnamefont
  {Berridge}},\ }\href {https://books.google.co.in/books?id=dKk0I-KMDJIC}
  {\emph {\bibinfo {title} {Biochemical Oscillations and Cellular Rhythms: The
  Molecular Bases of Periodic and Chaotic Behaviour}}}\ (\bibinfo  {publisher}
  {Cambridge University Press},\ \bibinfo {year} {1997})\BibitemShut {NoStop}%
\bibitem [{\citenamefont {Kar}\ and\ \citenamefont
  {Ray}(2003)}]{kar_2003_collapse}%
  \BibitemOpen
  \bibfield  {author} {\bibinfo {author} {\bibfnamefont {S.}~\bibnamefont
  {Kar}}\ and\ \bibinfo {author} {\bibfnamefont {D.~S.}\ \bibnamefont {Ray}},\
  }\href {\doibase 10.1103/PhysRevLett.90.238102} {\bibfield  {journal}
  {\bibinfo  {journal} {Phys. Rev. Lett.}\ }\textbf {\bibinfo {volume} {90}},\
  \bibinfo {pages} {238102} (\bibinfo {year} {2003})}\BibitemShut {NoStop}%
\bibitem [{\citenamefont {Ghosh}\ and\ \citenamefont
  {Ray}(2014)}]{ghosh_2014_lienard}%
  \BibitemOpen
  \bibfield  {author} {\bibinfo {author} {\bibfnamefont {S.}~\bibnamefont
  {Ghosh}}\ and\ \bibinfo {author} {\bibfnamefont {D.~S.}\ \bibnamefont
  {Ray}},\ }\href {\doibase 10.1140/epjb/e2014-41070-1} {\bibfield  {journal}
  {\bibinfo  {journal} {The European Physical Journal B}\ }\textbf {\bibinfo
  {volume} {87}},\ \bibinfo {pages} {65} (\bibinfo {year} {2014})}\BibitemShut
  {NoStop}%
\bibitem [{\citenamefont {Ghosh}\ and\ \citenamefont
  {Ray}(2015)}]{ghosh_2015_rayleigh}%
  \BibitemOpen
  \bibfield  {author} {\bibinfo {author} {\bibfnamefont {S.}~\bibnamefont
  {Ghosh}}\ and\ \bibinfo {author} {\bibfnamefont {D.~S.}\ \bibnamefont
  {Ray}},\ }\href {\doibase 10.1063/1.4931401} {\bibfield  {journal} {\bibinfo
  {journal} {The Journal of Chemical Physics}\ }\textbf {\bibinfo {volume}
  {143}},\ \bibinfo {pages} {124901} (\bibinfo {year} {2015})}\BibitemShut
  {NoStop}%
\bibitem [{\citenamefont {Linge}\ and\ \citenamefont
  {Langtangen}(2016)}]{linge_2016_programming}%
  \BibitemOpen
  \bibfield  {author} {\bibinfo {author} {\bibfnamefont {S.}~\bibnamefont
  {Linge}}\ and\ \bibinfo {author} {\bibfnamefont {H.}~\bibnamefont
  {Langtangen}},\ }\href {https://books.google.co.in/books?id=INlCDwAAQBAJ}
  {\emph {\bibinfo {title} {Programming for Computations - Python: A Gentle
  Introduction to Numerical Simulations with Python}}},\ Texts in Computational
  Science and Engineering\ (\bibinfo  {publisher} {Springer International
  Publishing},\ \bibinfo {year} {2016})\BibitemShut {NoStop}%
\bibitem [{\citenamefont {Jung}\ and\ \citenamefont
  {H\"anggi}(1990)}]{jung_1990_invariant}%
  \BibitemOpen
  \bibfield  {author} {\bibinfo {author} {\bibfnamefont {P.}~\bibnamefont
  {Jung}}\ and\ \bibinfo {author} {\bibfnamefont {P.}~\bibnamefont
  {H\"anggi}},\ }\href {\doibase 10.1103/PhysRevLett.65.3365} {\bibfield
  {journal} {\bibinfo  {journal} {Phys. Rev. Lett.}\ }\textbf {\bibinfo
  {volume} {65}},\ \bibinfo {pages} {3365} (\bibinfo {year}
  {1990})}\BibitemShut {NoStop}%
\bibitem [{\citenamefont {Barik}\ and\ \citenamefont
  {Ray}(2005)}]{barik_2005_quantum_2}%
  \BibitemOpen
  \bibfield  {author} {\bibinfo {author} {\bibfnamefont {D.}~\bibnamefont
  {Barik}}\ and\ \bibinfo {author} {\bibfnamefont {D.~S.}\ \bibnamefont
  {Ray}},\ }\href {\doibase 10.1007/s10955-005-5251-y} {\bibfield  {journal}
  {\bibinfo  {journal} {Journal of Statistical Physics}\ }\textbf {\bibinfo
  {volume} {120}},\ \bibinfo {pages} {339} (\bibinfo {year}
  {2005})}\BibitemShut {NoStop}%
\bibitem [{\citenamefont {Pikovsky}\ \emph {et~al.}(2001)\citenamefont
  {Pikovsky}, \citenamefont {Rosenblum},\ and\ \citenamefont
  {Kurths}}]{pikovskybook}%
  \BibitemOpen
  \bibfield  {author} {\bibinfo {author} {\bibfnamefont {A.~S.}\ \bibnamefont
  {Pikovsky}}, \bibinfo {author} {\bibfnamefont {M.}~\bibnamefont {Rosenblum}},
  \ and\ \bibinfo {author} {\bibfnamefont {J.}~\bibnamefont {Kurths}},\
  }\href@noop {} {\emph {\bibinfo {title} {Synchronization: A Universal Concept
  in Nonlinear Science}}}\ (\bibinfo  {publisher} {Cambridge University
  Press},\ \bibinfo {address} {New York},\ \bibinfo {year} {2001})\BibitemShut
  {NoStop}%
\bibitem [{\citenamefont {B.van~der Pol}(1928)}]{b_van_1928}%
  \BibitemOpen
  \bibfield  {author} {\bibinfo {author} {\bibfnamefont {J.~d.~M.}\
  \bibnamefont {B.van~der Pol}},\ }\href@noop {} {\bibfield  {journal}
  {\bibinfo  {journal} {Phil. Mag. Suppl.}\ }\textbf {\bibinfo {volume} {6}},\
  \bibinfo {pages} {763} (\bibinfo {year} {1928})}\BibitemShut {NoStop}%
\bibitem [{\citenamefont {R.Fitzhugh}(1961)}]{fitzhugh_1961}%
  \BibitemOpen
  \bibfield  {author} {\bibinfo {author} {\bibnamefont {R.Fitzhugh}},\
  }\href@noop {} {\bibfield  {journal} {\bibinfo  {journal} {Biophys.J}\
  }\textbf {\bibinfo {volume} {1}},\ \bibinfo {pages} {445} (\bibinfo {year}
  {1961})}\BibitemShut {NoStop}%
\bibitem [{\citenamefont {Jewett}\ and\ \citenamefont
  {Kronauer}(1998)}]{jewett_1998_refinement}%
  \BibitemOpen
  \bibfield  {author} {\bibinfo {author} {\bibfnamefont {M.~E.}\ \bibnamefont
  {Jewett}}\ and\ \bibinfo {author} {\bibfnamefont {R.~E.}\ \bibnamefont
  {Kronauer}},\ }\href {\doibase https://doi.org/10.1006/jtbi.1998.0667}
  {\bibfield  {journal} {\bibinfo  {journal} {Journal of Theoretical Biology}\
  }\textbf {\bibinfo {volume} {192}},\ \bibinfo {pages} {455 } (\bibinfo {year}
  {1998})}\BibitemShut {NoStop}%
\bibitem [{\citenamefont {Lee}\ and\ \citenamefont
  {Sadeghpour}(2013)}]{lee_2013_quantum}%
  \BibitemOpen
  \bibfield  {author} {\bibinfo {author} {\bibfnamefont {T.~E.}\ \bibnamefont
  {Lee}}\ and\ \bibinfo {author} {\bibfnamefont {H.~R.}\ \bibnamefont
  {Sadeghpour}},\ }\href {\doibase 10.1103/PhysRevLett.111.234101} {\bibfield
  {journal} {\bibinfo  {journal} {Phys. Rev. Lett.}\ }\textbf {\bibinfo
  {volume} {111}},\ \bibinfo {pages} {234101} (\bibinfo {year}
  {2013})}\BibitemShut {NoStop}%
\bibitem [{\citenamefont {Ishibashi}\ and\ \citenamefont
  {Kanamoto}(2017)}]{Ishibashi2017oscillation}%
  \BibitemOpen
  \bibfield  {author} {\bibinfo {author} {\bibfnamefont {K.}~\bibnamefont
  {Ishibashi}}\ and\ \bibinfo {author} {\bibfnamefont {R.}~\bibnamefont
  {Kanamoto}},\ }\href {\doibase 10.1103/PhysRevE.96.052210} {\bibfield
  {journal} {\bibinfo  {journal} {Phys. Rev. E}\ }\textbf {\bibinfo {volume}
  {96}},\ \bibinfo {pages} {052210} (\bibinfo {year} {2017})}\BibitemShut
  {NoStop}%
\bibitem [{\citenamefont {Walter}\ \emph {et~al.}(2014)\citenamefont {Walter},
  \citenamefont {Nunnenkamp},\ and\ \citenamefont
  {Bruder}}]{walter_2014_quantum}%
  \BibitemOpen
  \bibfield  {author} {\bibinfo {author} {\bibfnamefont {S.}~\bibnamefont
  {Walter}}, \bibinfo {author} {\bibfnamefont {A.}~\bibnamefont {Nunnenkamp}},
  \ and\ \bibinfo {author} {\bibfnamefont {C.}~\bibnamefont {Bruder}},\ }\href
  {\doibase 10.1103/PhysRevLett.112.094102} {\bibfield  {journal} {\bibinfo
  {journal} {Phys. Rev. Lett.}\ }\textbf {\bibinfo {volume} {112}},\ \bibinfo
  {pages} {094102} (\bibinfo {year} {2014})}\BibitemShut {NoStop}%
\bibitem [{\citenamefont {Lee}\ \emph {et~al.}(2014)\citenamefont {Lee},
  \citenamefont {Chan},\ and\ \citenamefont {Wang}}]{lee_2014_entanglement}%
  \BibitemOpen
  \bibfield  {author} {\bibinfo {author} {\bibfnamefont {T.~E.}\ \bibnamefont
  {Lee}}, \bibinfo {author} {\bibfnamefont {C.-K.}\ \bibnamefont {Chan}}, \
  and\ \bibinfo {author} {\bibfnamefont {S.}~\bibnamefont {Wang}},\ }\href
  {\doibase 10.1103/PhysRevE.89.022913} {\bibfield  {journal} {\bibinfo
  {journal} {Phys. Rev. E}\ }\textbf {\bibinfo {volume} {89}},\ \bibinfo
  {pages} {022913} (\bibinfo {year} {2014})}\BibitemShut {NoStop}%
\bibitem [{\citenamefont {Leibfried}\ \emph {et~al.}(2003)\citenamefont
  {Leibfried}, \citenamefont {Blatt}, \citenamefont {Monroe},\ and\
  \citenamefont {Wineland}}]{leibfried_2003_quantum}%
  \BibitemOpen
  \bibfield  {author} {\bibinfo {author} {\bibfnamefont {D.}~\bibnamefont
  {Leibfried}}, \bibinfo {author} {\bibfnamefont {R.}~\bibnamefont {Blatt}},
  \bibinfo {author} {\bibfnamefont {C.}~\bibnamefont {Monroe}}, \ and\ \bibinfo
  {author} {\bibfnamefont {D.}~\bibnamefont {Wineland}},\ }\href {\doibase
  10.1103/RevModPhys.75.281} {\bibfield  {journal} {\bibinfo  {journal} {Rev.
  Mod. Phys.}\ }\textbf {\bibinfo {volume} {75}},\ \bibinfo {pages} {281}
  (\bibinfo {year} {2003})}\BibitemShut {NoStop}%
\bibitem [{\citenamefont {Häffner}\ \emph {et~al.}(2008)\citenamefont
  {Häffner}, \citenamefont {Roos},\ and\ \citenamefont
  {Blatt}}]{haffner_2008_quantum}%
  \BibitemOpen
  \bibfield  {author} {\bibinfo {author} {\bibfnamefont {H.}~\bibnamefont
  {Häffner}}, \bibinfo {author} {\bibfnamefont {C.}~\bibnamefont {Roos}}, \
  and\ \bibinfo {author} {\bibfnamefont {R.}~\bibnamefont {Blatt}},\ }\href
  {\doibase https://doi.org/10.1016/j.physrep.2008.09.003} {\bibfield
  {journal} {\bibinfo  {journal} {Physics Reports}\ }\textbf {\bibinfo {volume}
  {469}},\ \bibinfo {pages} {155 } (\bibinfo {year} {2008})}\BibitemShut
  {NoStop}%
\bibitem [{\citenamefont {Blatt}\ and\ \citenamefont
  {Roos}(2012)}]{blatt_2012_quantum}%
  \BibitemOpen
  \bibfield  {author} {\bibinfo {author} {\bibfnamefont {R.}~\bibnamefont
  {Blatt}}\ and\ \bibinfo {author} {\bibfnamefont {C.~F.}\ \bibnamefont
  {Roos}},\ }\href {\doibase 10.1038/nphys2252} {\bibfield  {journal} {\bibinfo
   {journal} {Nature Physics}\ }\textbf {\bibinfo {volume} {8}},\ \bibinfo
  {pages} {277} (\bibinfo {year} {2012})}\BibitemShut {NoStop}%
\bibitem [{\citenamefont {Monroe}\ and\ \citenamefont
  {Kim}(2013)}]{monroe_2013_scaling}%
  \BibitemOpen
  \bibfield  {author} {\bibinfo {author} {\bibfnamefont {C.}~\bibnamefont
  {Monroe}}\ and\ \bibinfo {author} {\bibfnamefont {J.}~\bibnamefont {Kim}},\
  }\href {\doibase 10.1126/science.1231298} {\bibfield  {journal} {\bibinfo
  {journal} {Science}\ }\textbf {\bibinfo {volume} {339}},\ \bibinfo {pages}
  {1164} (\bibinfo {year} {2013})}\BibitemShut {NoStop}%
\bibitem [{\citenamefont {Rayleigh}(1883)}]{lord_1883_mainatained}%
  \BibitemOpen
  \bibfield  {author} {\bibinfo {author} {\bibfnamefont {L.}~\bibnamefont
  {Rayleigh}},\ }\href {\doibase 10.1080/14786448308627342} {\bibfield
  {journal} {\bibinfo  {journal} {Edinburgh, and Dublin Philosophical Magazine
  and Journal of Science}\ }\textbf {\bibinfo {volume} {15}},\ \bibinfo {pages}
  {229} (\bibinfo {year} {1883})}\BibitemShut {NoStop}%
\bibitem [{\citenamefont {Yariv}(2000)}]{yariv_2000_introduction}%
  \BibitemOpen
  \bibfield  {author} {\bibinfo {author} {\bibfnamefont {A.}~\bibnamefont
  {Yariv}},\ }\href {https://books.google.co.in/books?id=yFl8swEACAAJ} {\emph
  {\bibinfo {title} {Introduction to Optical Electronics}}}\ (\bibinfo {year}
  {2000})\BibitemShut {NoStop}%
\bibitem [{\citenamefont {Rychkov}(1975)}]{rychkov_1975_maximal}%
  \BibitemOpen
  \bibfield  {author} {\bibinfo {author} {\bibfnamefont {G.~S.}\ \bibnamefont
  {Rychkov}},\ }\href {http://www.ams.org/mathscinet-getitem?mr=369803}
  {\bibfield  {journal} {\bibinfo  {journal} {Differ. Uravn.}\ }\textbf
  {\bibinfo {volume} {11}},\ \bibinfo {pages} {390} (\bibinfo {year}
  {1975})}\BibitemShut {NoStop}%
\bibitem [{\citenamefont {Blows}\ and\ \citenamefont
  {Lloyd}(1984)}]{blows_1984_number}%
  \BibitemOpen
  \bibfield  {author} {\bibinfo {author} {\bibfnamefont {T.~R.}\ \bibnamefont
  {Blows}}\ and\ \bibinfo {author} {\bibfnamefont {N.~G.}\ \bibnamefont
  {Lloyd}},\ }\href {\doibase 10.1017/S0305004100061636} {\bibfield  {journal}
  {\bibinfo  {journal} {Mathematical Proceedings of the Cambridge Philosophical
  Society}\ }\textbf {\bibinfo {volume} {95}},\ \bibinfo {pages} {359–366}
  (\bibinfo {year} {1984})}\BibitemShut {NoStop}%
\bibitem [{\citenamefont {Giacomini}\ and\ \citenamefont
  {Neukirch}(1997)}]{giacomini_1997_number}%
  \BibitemOpen
  \bibfield  {author} {\bibinfo {author} {\bibfnamefont {H.}~\bibnamefont
  {Giacomini}}\ and\ \bibinfo {author} {\bibfnamefont {S.}~\bibnamefont
  {Neukirch}},\ }\href {\doibase 10.1103/PhysRevE.56.3809} {\bibfield
  {journal} {\bibinfo  {journal} {Phys. Rev. E}\ }\textbf {\bibinfo {volume}
  {56}},\ \bibinfo {pages} {3809} (\bibinfo {year} {1997})}\BibitemShut
  {NoStop}%
\bibitem [{\citenamefont {Llibre}\ \emph {et~al.}(1998)\citenamefont {Llibre},
  \citenamefont {Pizarro},\ and\ \citenamefont {Ponce}}]{libre_1998_limit}%
  \BibitemOpen
  \bibfield  {author} {\bibinfo {author} {\bibfnamefont {J.}~\bibnamefont
  {Llibre}}, \bibinfo {author} {\bibfnamefont {L.}~\bibnamefont {Pizarro}}, \
  and\ \bibinfo {author} {\bibfnamefont {E.}~\bibnamefont {Ponce}},\ }\href
  {\doibase 10.1103/PhysRevE.58.5185} {\bibfield  {journal} {\bibinfo
  {journal} {Phys. Rev. E}\ }\textbf {\bibinfo {volume} {58}},\ \bibinfo
  {pages} {5185} (\bibinfo {year} {1998})}\BibitemShut {NoStop}%
\bibitem [{\citenamefont {Ghosh}\ and\ \citenamefont
  {Ray}(2013)}]{ghosh_2013_chemical}%
  \BibitemOpen
  \bibfield  {author} {\bibinfo {author} {\bibfnamefont {S.}~\bibnamefont
  {Ghosh}}\ and\ \bibinfo {author} {\bibfnamefont {D.~S.}\ \bibnamefont
  {Ray}},\ }\href {\doibase 10.1063/1.4826169} {\bibfield  {journal} {\bibinfo
  {journal} {The Journal of Chemical Physics}\ }\textbf {\bibinfo {volume}
  {139}},\ \bibinfo {pages} {164112} (\bibinfo {year} {2013})}\BibitemShut
  {NoStop}%
\bibitem [{\citenamefont {Gaiko}(2008)}]{gaiko_2008_limit}%
  \BibitemOpen
  \bibfield  {author} {\bibinfo {author} {\bibfnamefont {V.}~\bibnamefont
  {Gaiko}},\ }\href@noop {} {\bibfield  {journal} {\bibinfo  {journal} {Cubo}\
  }\textbf {\bibinfo {volume} {10}},\ \bibinfo {pages} {115} (\bibinfo {year}
  {2008})}\BibitemShut {NoStop}%
\end{thebibliography}
\end{document}